\documentclass[sigconf]{acmart}

\AtBeginDocument{%
  \providecommand\BibTeX{{%
    Bib\TeX}}}

\copyrightyear{2025}
\acmYear{2025}
\setcopyright{rightsretained}
\acmConference[ASIA CCS '25]{ACM Asia Conference on Computer and Communications Security}{August 25--29, 2025}{Hanoi, Vietnam}
\acmBooktitle{ACM Asia Conference on Computer and Communications Security (ASIA CCS '25), August 25--29, 2025, Hanoi, Vietnam}\acmDOI{10.1145/3708821.3710832}
\acmISBN{979-8-4007-1410-8/25/08}

\usepackage{url}
\usepackage{accents}
\usepackage[linesnumbered,ruled]{algorithm2e}
\usepackage[noend]{algpseudocode}
\usepackage{booktabs}
\usepackage{graphicx}
\usepackage{textcomp}
\usepackage{xcolor}
\definecolor{Mahogany}{rgb}{0.75, 0.25, 0.0}
\usepackage{comment}
\usepackage{multicol,multirow}
\usepackage[T1]{fontenc}
\def\BibTeX{{\rm B\kern-.05em{\sc i\kern-.025em b}\kern-.08em
    T\kern-.1667em\lower.7ex\hbox{E}\kern-.125emX}}
\usepackage{todonotes}
\newsavebox{\measurebox}
\usepackage{threeparttable}
\usepackage[font=small]{caption}
\usepackage{subcaption}
\usepackage{bbding}
\usepackage{tikz}

\usepackage{dblfloatfix}

\newcommand{\homa}[1]{{{\color{orange}#1}}{}}

\makeatletter
\newcommand{\algorithmfootnote}[2][\footnotesize]{%
  \let\old@algocf@finish\@algocf@finish
  \def\@algocf@finish{\old@algocf@finish
    \leavevmode\rlap{\begin{minipage}{\linewidth}
    #1#2
    \end{minipage}}%
  }%
}
\makeatother

\newcommand*{\Scale}[2][4]{\scalebox{#1}{$#2$}}%

\newcommand*{\vsepfbox}[1]{%
  \begingroup
    \sbox0{\fbox{#1}}%
    \setlength{\fboxrule}{0pt}%
    \mbox{\kern-\fboxsep\fbox{\unhbox0}\kern-\fboxsep}%
  \endgroup
}

\begin{document}

\title{Runtime Stealthy Perception Attacks against DNN-based Adaptive Cruise Control Systems 
}




\author{Xugui Zhou}
\email{xuguizhou@lsu.edu}
\orcid{0000-0002-3663-7447}
\affiliation{%
  \country{Louisiana State University, USA}
}

\author{Anqi Chen}
\email{chen.anqi3@northeastern.edu}
\affiliation{%
  \country{Northeastern University, USA}
  }

\author{Maxfield Kouzel}
\author{Haotian Ren}
\affiliation{%
  \country{University of Virginia, USA}
}


\author{Morgan McCarty}
\email{morgannmccarty@gmail.com}
\affiliation{%
  \country{Northeastern University, USA}
  }

\author{Cristina Nita-Rotaru}
\email{c.nitarotaru@northeastern.edu}
\affiliation{%
  \country{Northeastern University, USA}
  }

\author{Homa Alemzadeh}
\email{ha4d@virginia.edu}
\affiliation{%
  \country{University of Virginia, USA}
}

\renewcommand{\shortauthors}{Zhou et al.}





\begin{abstract} 
%
Adaptive Cruise Control (ACC) is a widely used driver assistance technology for maintaining desired speed and safe distance to the leading vehicle. This paper evaluates the security of the deep neural network (DNN) based ACC systems under runtime stealthy perception attacks that strategically inject perturbations into camera data to cause forward collisions. We present a context-aware strategy for the selection of the most critical times for triggering the attacks and a novel optimization-based method for the adaptive generation of image perturbations at runtime. 
We evaluate the effectiveness of the proposed attack using an actual vehicle, a publicly available driving dataset, and a realistic simulation platform with the control software from a production ACC system, a physical-world driving simulator, and interventions by the human driver and safety features such as Advanced Emergency Braking System (AEBS). 
Experimental results show that the proposed attack achieves 142.9 times higher success rate in causing hazards and 82.6\% higher evasion rate than baselines, while being stealthy and robust to real-world factors and dynamic changes in the environment.
This study highlights the role of human drivers and basic safety mechanisms in preventing attacks.

\end{abstract}



\maketitle

\section{Introduction}
\label{sec:intro}
Level-2 Advanced Driver Assistance Systems (ADAS) 
provide autonomous driving features while still requiring human
attention at all times \cite{SAEroadmap}. Examples include Adaptive Cruise Control (ACC) which controls longitudinal movement, Automatic Lane Centering (ALC) which controls lateral movement, and Advanced Emergency Braking System (AEBS) which controls braking through Automatic Emergency Braking (AEB) and provides warnings through Forward Collision Warning (FCW). Over 17 million passenger cars worldwide are equipped with ADAS \cite{canalys1}.

One important ADAS feature is ACC, which makes highway driving more comfortable by automatically changing the speed when traffic slows down or speeds up. ACC takes as input sensor measurements such as radar, Lidar, or camera and adjusts the speed to maintain a safe following distance to the lead vehicle \cite{sayer2003adaptive,sae2018taxonomy}. 
At the core of ACC lies the detection and tracking of the lead vehicle. Highly accurate methods \cite{ravindran2020multi, sharma2018beyond} for detection and tracking rely on Deep Learning (DL) based object detection using camera or fusion of camera and radar/Lidar data. A Longitudinal planner (LP) uses the prediction from the DL module to compute the desired speed and acceleration. Malfunctioning of the object detection module can have serious consequences including accidents. Given the critical role of object detection and tracking in the safety of ACC and that many commercial ACC systems (e.g., Tesla Autopilot \cite{autopilot}, Comma.ai OpenPilot \cite{commaai-openpilot}) use DL-based object detection,
these mechanisms must function correctly under any conditions, including in the presence of adversaries.

Previous work has shown attacks against Deep Neural Networks (DNN) used for perception such as adversarial perturbations \cite{zhou2022adversarial}, adversarial patches \cite{liu2018dpatch,lee2019physical,sato2021dirty} or well-crafted stickers on road signs \cite{eykholt2018robust}, the road \cite{tencent2019experimental}, or camera lenses \cite{li2019adversarial} to
change the predicted class or probability of detecting a target object or the lane lines. However, misprediction of the lane lines only affects the lateral control (ALC) and
simply changing the lead object class or detection probability does not necessarily impact the LP enough to cause unsafe ACC behavior (e.g., sudden acceleration).  Attacking DNN-based ACC systems necessitates influencing the relative distance and speed with respect to the lead vehicle. One such attack using physical adversarial patches to control the relative distance and speed to the lead vehicle was shown against a production ACC~\cite{guo2023adversarial}. However, the attack required placing a conspicuous large patch on the back of a truck and driving the truck in front of the target vehicle, a method easily noticeable or preventable by human drivers (see Appendix \ref{sec:appendix:userstudy}). 

In terms of computational effort, the perturbation-based attacks described above
predominantly rely on \textit{offline} optimizations.
For attacks on ACC to be effective and robust, perturbations must be created at \textit{runtime} by considering dynamic factors such as the lead vehicle's size and position across consecutive camera frames, inter-dependencies across the frames, and environmental changes, while satisfying runtime computation constraints.

Notably, previous work ignored the presence of safety interventions in the ADAS control loop. Although some attack works have considered AEB or FCW, they either do not apply to ACC \cite{sato2021dirty} or directly change the state estimations \cite{ma2021sequential}. Some recent works \cite{zhou2022strategic, jha2020ml} have used the system context to determine the optimum time for attack injection. However, they did not directly compute the perturbations for the DL perception module to affect relative speed and distance and could be easily detected by existing safety mechanisms or anomaly detection methods \cite{zhou2023hybrid, Choi2020software}.

To fill these gaps, in this paper, 
we focus on runtime stealthy attacks against DNN-based ACC systems that inject minimal image perturbations into the DNN input with the goal of causing ACC controller to issue unsafe acceleration commands that cannot be mitigated by the human driver or existing safety mechanisms and lead to safety hazards, such as forward collision. 
We assume baseline ACC, ADAS software, and existing safety mechanisms are trusted.

Designing  \textit{stealthy safety-critical attacks} in the \textit{human-in-the-loop ADAS} is a challenging task as the attackers need to explore the extensive attack parameter space to devise a strategy for effectively manipulating the DNN inputs and causing unsafe driving behavior while considering the dynamic changes of the environment at runtime, real-time constraints, and safety interventions. 
We propose a \textit{context-aware} attack strategy {and \textit{optimization} method} together with a \textit{safety intervention} simulator to explore key questions on how the \textit{timing} and \textit{value} of perturbations affect the success of attacks in (1) causing safety hazards and (2) evading human driver or safety mechanism detection and intervention. 
To the best of our knowledge, this is the first evaluation of the security of DNN-based production ACC systems under \textit{runtime strategic attacks} using a \textit{combined knowledge-and-data-driven} approach by taking into account \textit{human drivers} and \textit{realistic safety mechanisms}. This study provides insights into the vulnerabilities and risks associated with DNN-based ACC systems and the role of human operators and safety mechanisms in preventing attacks. 

The main contributions of the paper are the following:
\vspace{-1em}
\begin{itemize}


\item We adopt a control-theoretic hazard analysis method to identify the most critical system contexts for launching attacks that maximize the chance of forward collisions. 

\item We design a novel optimization-based approach and an adaptive algorithm to generate stealthy image perturbations and add them in the form of an adversarial patch to the input camera frames at \textit{runtime} to fool DNN model and cause unsafe acceleration by ACC controller before being detected or mitigated by the human driver or ADAS safety mechanisms. 


\item We evaluate the effectiveness of the attacks with a real vehicle and driving dataset and a realistic simulation platform that integrates an open-source ADAS control software, OpenPilot from Comma.ai (with over 10,000 active users on the road)~\cite{commaai-openpilot} and the state-of-the-art physical-world driving simulator, CARLA, with a driver reaction simulator and the typical ADAS safety mechanisms (AEB and FCW), which we implement based on the international standards. 

\item Experimental results show that our context-aware attack strategy causes 28.6x more hazards than random attacks.
The proposed optimization-based perturbation algorithm achieves a 100\% attack success rate in four high-risk driving scenarios (in the absence of safety interventions), 142.9x and 1.9x higher than random value perturbation and APGD-based methods \cite{croce2020reliable}. 
Our approach is also stealthy and robust to real-world factors such as different camera positions, distances to lead vehicle, and weather and lighting conditions.

\item 
We observe similar results in the presence of the human driver and safety feature interventions, where our attack still achieves an 82.6\% success rate, while all the random and APGD-based attacks are mitigated by the safety mechanisms.  


\end{itemize}



\textbf{Ethics.} We have submitted responsible disclosures to Comma.ai. For the human subject study, we received IRB approval and followed the IRB requirements for the recruitment of participants and conduct of experiments.

\section{ADAS Overview}
\begin{figure}[b!]
    \centering
    \vspace{-1em}
    \includegraphics[width=\columnwidth]{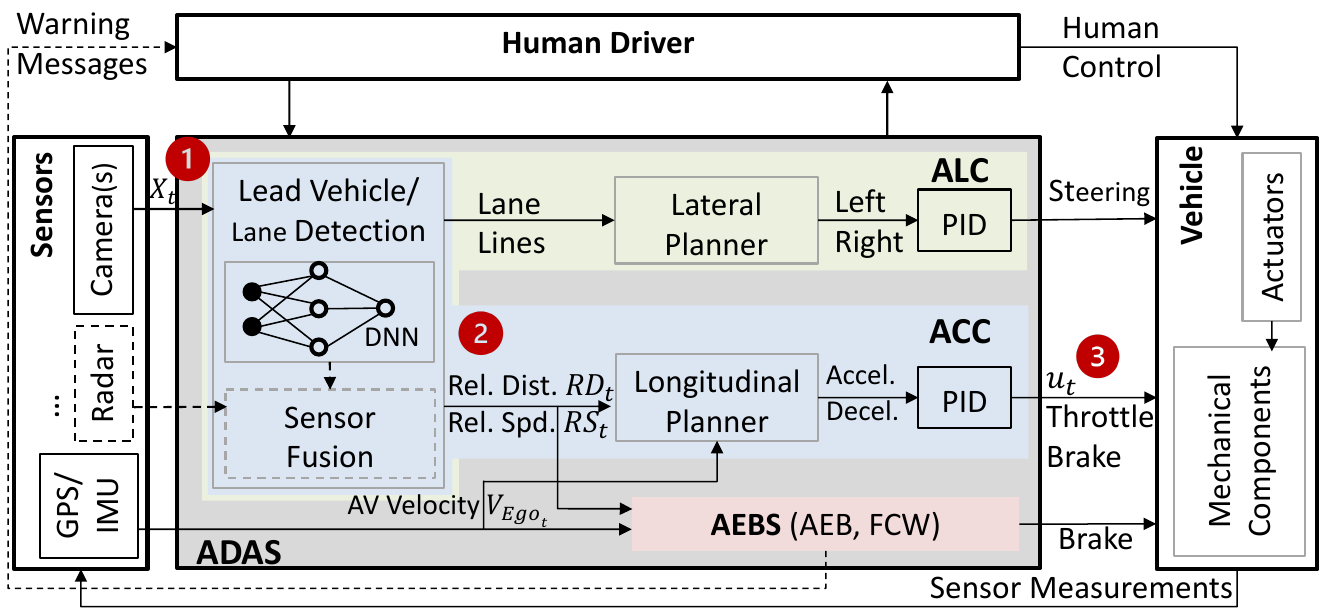}
    \vspace{-2em}
    \caption{ADAS architecture with ACC, AEBS, and ALC.
    }
    \label{fig:AdasOverview}
\end{figure}

Fig. \ref{fig:AdasOverview} shows the overall structure of a typical ADAS, including ACC, AEBS, and ALC features.

\subsection{Adaptive Cruise Control (ACC)}
\label{sec:ACC}
The main goal of ACC is to {maintain a safe following distance between the autonomous vehicle (referred to as Ego vehicle or AV) and the vehicle driving in the same lane in front of the AV (referred to as lead vehicle or LV) by adjusting the AV speed based on the {estimated relative distance and relative speed to the LV.} 
}


\textbf{Sensors.}
Existing DNN-based ACC systems either use camera data (e.g., Tesla Autopilot, Subaru Eyesight) or both camera and radar data (e.g., Apollo \cite{Apollo} and OpenPilot) to predict and track LVs and objects. 
Other sensors, such as GPS or IMU, are also used to detect current speed to match the target speed set by the drivers. 


\textbf{Lead Vehicle Detection.} 
The most critical part of ACC is lead vehicle detection (LVD), which {estimates the relative speed ($RS$) and distance ($RD$)} to the LV using camera data or a fusion of camera and radar data. 
Sensor fusion is the process of combining measurements from multiple sensors (e.g., camera and radar) usually using a Kalman filter \cite{bishop2001kalman} to overcome the limitations of individual sensors and obtain a more accurate perception of the surrounding environment.
Based on the LVD outputs, the main driving control actions (i.e., acceleration, deceleration, braking) are determined. 

\textbf{Longitudinal Planner.}
The next stage involves determining the optimal speed based on the LVD outputs and current vehicle state. The longitudinal planner uses algorithms such as Model Predictive Control (MPC) to generate multiple desired speed trajectories, each representing a series of speeds in a certain following period~\cite{camacho2013model}.

\textbf{Vehicle Control.}
At each control cycle $t$, the plan from the longitudinal planner with the lowest speed and risk (e.g., risk of colliding with the LV) is selected by the ACC system and fed to a Propotional-Integral-Derivative (PID) \cite{bishop2011modern} controller to get the specific optimal control command $u_t$ in the form of the throttle or brake amount such that the vehicle accurately and quickly follows the desired speed trajectory. 
Upon execution of the control command by the actuators, the vehicle's physical state, $s_t$ (e.g., current speed, location), transitions to a new state $s_{t+1}$.




\begin{table}[t!]
\footnotesize
\centering
\caption{Comparison of attacks on {DNN-based} ADAS.}
\vspace{-1.5em}
\label{tab:comparison}
\resizebox{\columnwidth}{!}
{%
\small
\begin{threeparttable}
    
\begin{tabular}{lllllll}
\toprule
\textbf{Attack} &   \multicolumn{3}{c}{\textbf{Attack} }& \multicolumn{2}{c}{\textbf{Safety Interventions}} & \textbf{Autonomy} \\  \cmidrule(lr){2-4}  \cmidrule(lr){5-6}
 \textbf{Method}& \textbf{Type}& \textbf{Vector}&\textbf{Target}& \textbf{AEBS} & \textbf{Driver} &\textbf{Level}\cite{SAEroadmap}\\
\midrule
\cite{eykholt2018robust}  & \multirow{4}{*}{\begin{tabular}[c]{@{}l@{}}Offline\end{tabular}}& Stickers on road signs  &Classifier & N & N &N/A\\
\cite{ma_wip_2023}        && Patch projected         &MTO& N & N & L4\\
\cite{tencent2019experimental} && Stickers on road  &ALC & N & N &L2\\
\cite{guo2023adversarial} && Patch on truck  &ACC & N & N & L2\\ 
 \cite{sato2021dirty} & & Patch on road  & ALC & \textbf{Y} & N &L2\\ \midrule

\cite{jia2019fooling} & \multirow{4}{*}{Runtime}& Perception inputs  &MTO & N & N & L4\\
\cite{jha2020ml}  & & Perception inputs  &MTO & No FCW & N &L4\\
\cite{ma2021sequential} & & Inner state variables &FCW & No AEB & Y &N/A\\
\cite{zhou2022strategic}  & & Control commands  &ACC, ALC & No AEB& \textbf{Y} &L2\\
Ours  & & Perception inputs  &ACC & \textbf{Y} & \textbf{Y} &L2\\
\bottomrule

\end{tabular}%

\begin{tablenotes}
    \item[*]MTO: Multi-Object Tracking;
\end{tablenotes}

\end{threeparttable}

}
\vspace{-2em}
\end{table}

\subsection{ADAS Safety Mechanisms}
\label{sec:bkground:safety}
The Advanced Emergency Braking System (AEBS), including FCW and AEB, is a fundamental ADAS safety mechanism that alerts drivers about potential collision risks with a lead obstacle and actively decelerates the vehicle to prevent accidents. 
As shown in Fig. \ref{fig:AdasOverview}, most AEBS implementations utilize both camera and radar for collision prediction through sensor fusion \cite{AdasSensorFusion} and make control actions based on the LVD outputs and other sensor measurements (see Appendix \ref{appendix:survey-aebs}).
In addition, some safety principles (such as maximum acceleration limits as required by international
standards, e.g., ISO 22179) or firmware safety checks (e.g., constraints on the output steering angle) are incorporated into the design of typical production ADAS to ensure driving safety \cite{openpilot_safety}. 

Previous studies on security of autonmous driving either focus on Level 4 or fully autonomous vehicles without  considering the impact of the human driver interventions during an emergency situation (e.g., abnormal acceleration)
or have overlooked the inclusion of basic safety features like AEB or FCW and their impact on the attack effectiveness (see Table \ref{tab:comparison}).
For a realistic assessment of ACC security, it is essential to evaluate the interventions of these safety features. 
At Level 2, drivers must maintain control and supervise ADAS functionalities \cite{SAEroadmap}. 
There exists a research gap concerning how to make the combination of a human driver and an autonomous vehicle acceptably safe. 
The primary challenge lies in assessing the ability of human drivers to anticipate and respond to situations where automation may fail. 

\vspace{-0.5em}
\subsection{OpenPilot}
\label{sec:background:platform}
We use a production ADAS called OpenPilot from Comma.ai \cite{commaai-openpilot} as our case study. OpenPilot is the only open-source production Level-2 ADAS, designed with the goal of improving visual perception and automated control (with ACC and ALC) through installing custom hardware to the OBD-II port on a vehicle. The targeted ACC system in OpenPilot follows the typical DNN-based ACC system architecture described in Fig. \ref{fig:AdasOverview} with an end-to-end system design \cite{chen2022level}.  Currently, OpenPilot supports over 250 car models (e.g., Toyota, Honda, etc.) \cite{Carmodels-OpenPilot}, has more than 10,000 active users and has accumulated a total driving distance of over 100 million miles on actual roads \cite{commaai-openpilot}. It is reported to achieve state-of-the-art autonomous driving performance, beating 17 existing production ADAS on the market 
in overall ranking by Consumer Reports \cite{CR-OpenPilot}. 

The DNN model used by OpenPilot, called Supercombo, utilizes an EfficientNet-B2 based CNN model to process image data \cite{schmedding2024strategic}. It incorporates the state of the vehicle and the environment by adding additional inputs from traffic conventions and the desired state. Multiple branches of GEMM (General Matrix Multiply) operations are then used to derive various predictions, such as lane lines, LVs, and vehicle pose, resulting in a total of 6,472 outputs \cite{Supercombo}. 

Comma.ai offers the community a closed-loop simulation environment \cite{commaai-openpilot} that integrates OpenPilot with a physical-world open urban driving simulator called CARLA, which can generate near-real high-quality camera image frames of the environment and has been widely used in the literature of autonomous driving \cite{Dosovitskiy17}. 
However, in this default simulation, the sensor fusion only relies on the camera data since no radar sensor is available. Further, none of the typical ADAS safety mechanisms are included.
\section{Runtime Context-Aware Perception Attack}
\label{sec:attackdesign}
In this section, we introduce our attack model, attack challenges, and runtime stealthy perception attack design.


\subsection{Attack Model}
\label{sec:attackmodel}

\textbf{Attacker Objective.}
The objective of the attacker is to maximize the error in LV predictions by the LVD's DNN module and cause forward collisions, while remaining stealthy to avoid being detected or prevented by driver or safety mechanisms (e.g., AEBS).

The attack is crafted in a stealthy way such that it is not distinguishable from noise, human errors, or accidents. This enables it to remain hidden longer and makes it less easily detected{/prevented} by existing defense mechanisms {(e.g., anomaly detection \cite{AVattacksurvey2021} or input transformation \cite{zhang2019defending,dziugaite2016study,xu2017feature,xu2017feature})}. 
The attacker can accomplish this by targeting the DNN inputs (\tikz[baseline=(char.base)]{\node[shape=circle,draw=Mahogany,fill=Mahogany,text=white,inner sep=1pt] (char) {1}} in Fig. \ref{fig:AdasOverview}) or directly manipulating the DNN outputs (\tikz[baseline=(char.base)]{\node[shape=circle,draw=Mahogany,fill=Mahogany,text=white,inner sep=1pt] (char) {2}}), depending on their capabilities and access level to the ACC system. In this paper, we primarily focus on DNN inputs to enhance stealthiness, as detailed in Section \ref{sec:perception_vs_output}.

\textbf{Attacker Knowledge.} 
We assume the attacker gains comprehensive knowledge about the target ACC system design and implementation by reverse engineering a purchased or rented vehicle with identical control software as the victim vehicle \cite{tencent2019experimental, AVattacksurvey2021} or by studying publicly available documents or source code. This is possible given some production ACC systems are open source \cite{commaai-openpilot,Apollo}.

\textbf{Attacker Capabilities.}
%
We assume the attacker has the capability to \textit{intercept {and change} live camera image frames} at runtime to compute an adversarial patch and fool the DNN model of ACC.  

A possible way to achieve this is to implant malware by
compromising the over-the-air (OTA) update mechanisms \cite{Cybersecurity,wen2020plug,mocnik_vehicular_nodate,elkhail2021vehicle} or gaining one-time remote access to the ADAS software through scanning the network, accessing stolen credentials and exploiting the vulnerabilities in SSH protocol~\cite{vulssh}, browsers, access control \cite{nie2017free}, wireless communications \cite{nie2017free, luo2022risks, hackjeep2015}, third-party components connected to in-vehicular network \cite{koscher2010experimental}, or some remote service/backdoor offered by the manufacturer (e.g., Comma Connect for OpenPilot \cite{comma-connect} or Bluelink for Hyundai). For example, a publicly-available tool developed for OpenPilot enables an attacker on the same network as a target device to install a malicious code~\cite{Workbench}. 
With such remote access, the attacker can also change the OTA settings (e.g., remote URL) to prevent potential patches from being effective. 
This assumption about the attack surface for deploying malware is also supported by previous works \cite{elkhail2021vehicle,eiza2017driving}, and could have a large impact as it can be generalized to any vehicle with similar OTA and DNN mechanisms and target a large fleet of vehicles at the same time.



\begin{table}[t!]
\centering
\caption{Threat models: attacker strength, capability, and impact.}
\vspace{-1em}
\label{tab:threatmodel}
\resizebox{\columnwidth}{!}{%
\begin{threeparttable}
\begin{tabular}{cccccll}
\toprule
\multirow{2}{*}{\textbf{\begin{tabular}[c]{@{}c@{}}Threat \\ Model\end{tabular}}} & \multirow{2}{*}{\textbf{\begin{tabular}[c]{@{}c@{}}Attacker \\ Strength\end{tabular}}} & \multirow{2}{*}{\textbf{\begin{tabular}[c]{@{}c@{}}Access to\\ ADAS Software\end{tabular}}} & \multirow{2}{*}{\textbf{\begin{tabular}[c]{@{}c@{}}Vehicular \\ Networks\end{tabular}}} & \multirow{2}{*}{\textbf{\begin{tabular}[c]{@{}c@{}}{Computation} \\ {Location}\end{tabular}}}&\multicolumn{1}{c}{\multirow{2}{*}{\textbf{Impact}}} & \multirow{2}{*}{\textbf{Examples}} \\ 
\multicolumn{1}{c}{} & \multicolumn{1}{c}{} &  &  &  &   &  \\ \midrule

Malware & Strong$^1$ & $\checkmark$ & r/w$^*$ & within ADAS& \begin{tabular}[c]{@{}l@{}}{Fleet of} \\ Vehicles\end{tabular} & \cite{elkhail2021vehicle,eiza2017driving} \\

Wireless & Medium$^2$  &  &r/w  & \begin{tabular}[c]{@{}l@{}} 
     Local Device,\\
     Remote Server
\end{tabular}  & \begin{tabular}[c]{@{}l@{}}Single \\ Vehicle\end{tabular} & \begin{tabular}[c]{@{}l@{}} \cite{lagraa2019real}\cite{hackcamera}\cite{jha2020ml} \\ \cite{nie2017free}\cite{miller2015remote} \end{tabular} \\

Physical & Weak$^3$ &  & r & Remote Server & \begin{tabular}[c]{@{}l@{}}Single \\ Vehicle\end{tabular} & \begin{tabular}[c]{@{}l@{}}\cite{hoory2020dynamic}\cite{chahe2023dynamic} \\\cite{lovisotto2021slap}\cite{man2023person}\cite{ma2023wip} \end{tabular} \\ \bottomrule
\end{tabular}%

\begin{tablenotes}
    \item[1] Other malware attacks are possible (e.g., DNN output, controller output);
    \item[2] Other sensor/actuator attacks are possible (e.g., RADAR, GPS, controller output);
    \item[3] Only perception attacks possible; 
    \item[*] r/w represents read (r) and write (w) access to vehicular networks.

\end{tablenotes}
\end{threeparttable}
}
\vspace{-2em}
\end{table}

Another way to compromise live camera data is to connect to a wireless communication device, either a third-party component or one implemented by an attacker, connected to the vehicular network, such as ROS communication channels \cite{lagraa2019real}, CAN Bus~\cite{nie2017free,openpilot_hardware,miller2015remote,wen2020plug} or Ethernet {channel} \cite{jha2020ml,hackcamera}), to read and send image data at runtime.
{The attacker computes the attack value on a local wireless device or a remote server.}

Further, physical attack methods are also viable, such as by displaying the patch on a monitor attached on the rear side of a leading adversarial vehicle \cite{hoory2020dynamic,chahe2023dynamic} or projecting the patch into the rear of the LV using a projector \cite{lovisotto2021slap,man2023person,ma2023wip}.


Table \ref{tab:threatmodel} summarizes various methods for runtime reading and modifying of live camera frames, given different attacker strengths and capabilities. {In this paper, we mainly focus on the runtime and optimized modification of live camera frames to enhance attack success rate and stealthiness, regardless of the threat model and how the attacker obtained access.} 
{In our experiments, we implement the attack through malicious OTA update to OpenPilot (Section {\ref{subsec:attackdesign}}).}

\textbf{Attacker Constraints.} 
We restrict the scope of attack capabilities in the paper to reading and modifying live camera data, ensuring uniformity across all presented threat models. Although some attack models could potentially enable more aggressive attacks, {such as directly altering ACC controller outputs (\tikz[baseline=(char.base)]{\node[shape=circle,draw=Mahogany,fill=Mahogany,text=white,inner sep=1pt] (char) {3}} in Fig. \ref{fig:AdasOverview}) via malware or wireless method or changing the DNN output (\tikz[baseline=(char.base)]{\node[shape=circle,draw=Mahogany,fill=Mahogany,text=white,inner sep=1pt] (char) {2}}) with malware method,}
these 
may be easier or earlier identified by safety mechanisms (e.g., AEBS) or human driver (see Section \ref{sec:perception_vs_output}).

%

The attacker does not consider injecting or replaying pre-recorded fake video frames as they need to be perfectly engineered offline or be pre-recorded, would not suit the constantly changing environment (e.g., surrounding vehicles, road conditions) at runtime, and could be easily noticed by human drivers (see Appendix \ref{sec:appendix:fakevideo}). 

\subsection{Attack Challenges}
\label{sec:challenges}


Several challenges need to be addressed in attacking DNN-based ACC systems at runtime.

\textbf{C1. Optimal timing of attacks at runtime to cause safety hazards.}
Prior attacks on ADAS that rely on random strategies to determine the attack timing (start time and duration) have proven ineffective in achieving a high attack success rate \cite{zhou2022strategic,rubaiyat2018experimental,jha2019ml} as they waste computational resources by trying random attack parameters that lead to no safety hazards. 
For instance, initiating an attack on an Ego vehicle to induce sudden acceleration does not cause safety hazards when no LV is detected.
Recent works have focused on using machine learning to explore the fault/attack parameter space \cite{moradi2020exploring} and improve the attack effectivenes~\cite{jha2019ml,jha2020ml}, but they still require substantial amounts of data from random attack experiments for model training.  
Finding the optimal triggering time and duration is crucial for effective attacks, yet challenging due to the vast parameter space that needs exploration.
\textbf{C2. Generating attack value at runtime to {adapt to} dynamic changes in the driving environment.}
Attacking DNN-based ACC systems on a moving vehicle faces challenges due to continuous variations in the driving environment, such as object position and size captured by the Ego vehicle's camera. Existing attack algorithms \cite{eykholt2018robust, guo2023adversarial} are inadequate as they plan perturbations offline, assuming fixed sizes and locations for attack vectors. A new algorithm is needed to dynamically adapt the attack vector's value (e.g., position, dimension, and amount of perturbation) to match the LV's dynamics. These changes disrupt the original attack vector generation process, requiring a unique approach to address inconsistencies and non-differentiability in the objective function.
In addition, the attack value should be designed in a stealthy way to avoid detection by the human driver or safety mechanisms.

%

\textbf{C3. Incorporating real-time constraints into the attack optimization process.}
Previous attacks on DNN models assume predetermined target images \cite{eykholt2018robust} or a known set \cite{sato2021dirty,guo2023adversarial} with unlimited computation resources, allowing iterative optimization until an optimal attack vector is generated. However, attacking ACC systems in real-time presents challenges as the camera continuously provides frames without prior knowledge. An attack vector must be generated in real-time before the next frame or control action execution. The real-time control cycle and camera update frequency limit the speed of generating the adversarial attack vector and the frequency of assessing the perturbation's impact on DNN predictions. These tight constraints in typical ACC systems (e.g., frame rate of 20Hz and control cycle of 10ms in \cite{commaai-openpilot}) significantly impact the effectiveness of optimization-based attack strategies.


\subsection{Attack Design}
\label{subsec:attackdesign}
\begin{figure}[t!]
    \centering
    \includegraphics[width=\columnwidth]{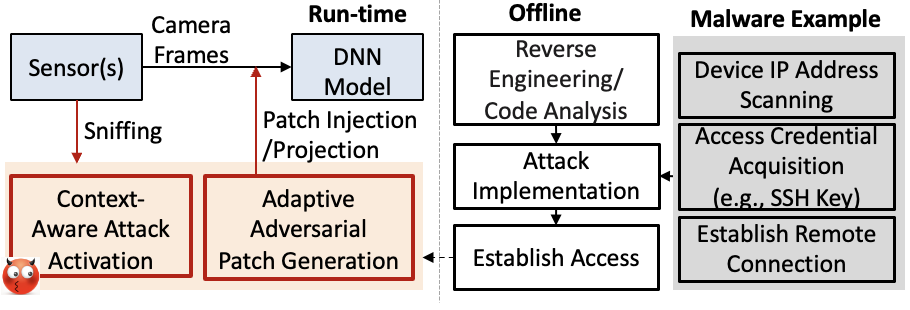}
    \vspace{-2.5em}
    \caption{Attack design: Offline preparation, Runtime execution.}
    \vspace{-1.5em}
    \label{fig:designoverview}
\end{figure}

Fig. \ref{fig:designoverview} illustrates the overall design of our attack, including the steps during the runtime execution and offline preparation. 

To tackle challenge \textbf{C1}, rather than randomly or exhaustively exploring the attack parameter space, 
we systematically characterize specific values within the parameter space (e.g., attack start time and duration). This targeted approach aims to identify optimal \textit{system contexts} (or critical times) for activating the attacks to not waste time and resources on non-hazardous scenarios. 

To address the challenges \textbf{C2-C3}, we design a novel optimization approach and an adaptive algorithm to dynamically determine optimal pixel values for an adversarial patch at runtime, aiming to maximize the error in DNN-based LV predictions and to accommodate dynamic changes in the driving environment. 
To optimize {attack effectiveness and computational efficiency}, our method focuses on the small area of the bounding box around the target vehicle and employs a primary attribution algorithm~\cite{sundararajan2017axiomatic} to identify and manipulate the most crucial pixels (Section \ref{sec:attackvaluedesign}). 
{Additionally, our adaptive algorithm retains optimization results of perturbation size, position, and value from the previous perception cycle instead of restarting the optimization process (contrary to other iterative optimization methods \cite{croce2020reliable}) to satisfy the tight real-time constraints of the perception system (50ms).}

The attacker initiates the attack process by conducting an offline analysis on the target DNN-based ACC system. 
The analysis includes examining operational data and open-source code to identify the target software files and functions and/or the DNN input and output fields to be monitored and infected. 
{Then the attack steps outlined in the orange box in Fig. \ref{fig:designoverview}, are implemented and executed either remotely or locally through unauthorized access to the target ACC under one {of} the threat models described in Table \ref{tab:threatmodel}. }
{In Fig \ref{fig:designoverview}, we present an attack implementation example via malware method} that will replace the target functions or system libraries. The attacker installs the malware on the target ACC system through one-time access to the victim vehicle control system, achieved by exploiting remote access vulnerabilities or compromising OTA updates (Section \ref{sec:attackmodel}). Appendix \ref{Appendix:deployment} provides an illustrative example of how we install the malware on an OpenPilot system.  At runtime, the malicious code intercepts sensor data before reaching the DNN model and infers the current system context for activating attacks. When the detected system context aligns with a critical system context (Section \ref{sec:CA}), an optimized adversarial patch is generated (Section \ref{subsec:objectivefunction}) and added to the image data and sent to DNN model. 

\subsubsection{Context-Aware Attack Activation} 
\label{sec:CA}
To find the most critical times for activation of the attack (\textbf{C1}), we adopt a control-theoretic hazard analysis method \cite{leveson2013stpa} to identify the most critical system contexts under which specific control actions are unsafe and, if issued by the ACC control software, could lead to hazards. 
This approach mainly relies on domain knowledge about system safety requirements and, contrary to an ML-based approach, does not require large amounts of training data or computation resources. 

In our hazard analysis, we define the accident as the adverse event of forward collision (measured by a zero or negative relative distance between the Ego vehicle and the LV or a front object). {This }can happen as a result of the Ego vehicle transitioning into a hazardous state that violates the safe following distance with LV.  

To determine the critical system contexts, we assess all the combinations of system states and ACC control actions (e.g., acceleration, deceleration) to identify the specific combinations that are most likely to lead to hazards. Table \ref{tab:contexttable} shows part of the safety context table for ACC with a focus on the acceleration commands. 
 Here, the critical system context is described as when the Headway Time (HWT, the time the Ego vehicle takes to drive the relative distance ($RD$) from the LV with the current speed) is less than a safety limit HWT$_{safe}$ (e.g., 2-3s), and the Ego vehicle is faster than the LV ($V_{Ego} > V_{Lead}$). Under such system context, an acceleration command induced by the camera perception attacks or other reasons will most likely lead to a forward collision hazard.
This high-level specification of critical system context can be done by an attacker based on the knowledge of the typical functionality of an ACC system and be applied to any ACC system with the same functional specification. 

\begin{table}[t!]
\centering
\caption{Partial safety context table for an ACC system.}
\vspace{-1em}
\label{tab:contexttable}
\resizebox{\columnwidth}{!}
{%
\begin{threeparttable}

\begin{tabular}{lllll}
\toprule
\textbf{Rule} &\multicolumn{2}{c}{\textbf{System Context}}                 & \textbf{Control Action}                                                                  & \textbf{Potential Hazards?} \\ \midrule
1 & \multirow{2}{*}{HWT$\leqslant$HWT$_{safe}$}    & RS$\leqslant$0    & \multirow{4}{*}{\begin{tabular}[c]{@{}l@{}}Acceleration   \end{tabular}} & No               \\
2&                                        & RS>0 &                                                                                  & Yes              \\ \cmidrule(l){1-3} \cmidrule(l){5-5}
3&\multirow{2}{*}{HWT>HWT$_{safe}$} & RS$\leqslant$0    &                                                                                  & No               \\
4&                                        & RS>0 &                                                                                  & No              \\\bottomrule
\end{tabular}

\begin{tablenotes}
\footnotesize
\item [*] HWT: Headway Time = Relative Distance/Current Speed; 
\item [*] RS: Relative Speed = Current Speed ($V_{Ego}$) - Lead Speed ($V_{Lead}$);

\end{tablenotes}

\end{threeparttable}
}
\vspace{-1em}

\end{table}

\subsubsection{Adaptive Adversarial Patch Generation}
\label{sec:attackvaluedesign}
\begin{figure}[t!]
    \centering
    \includegraphics[width=0.9\columnwidth]{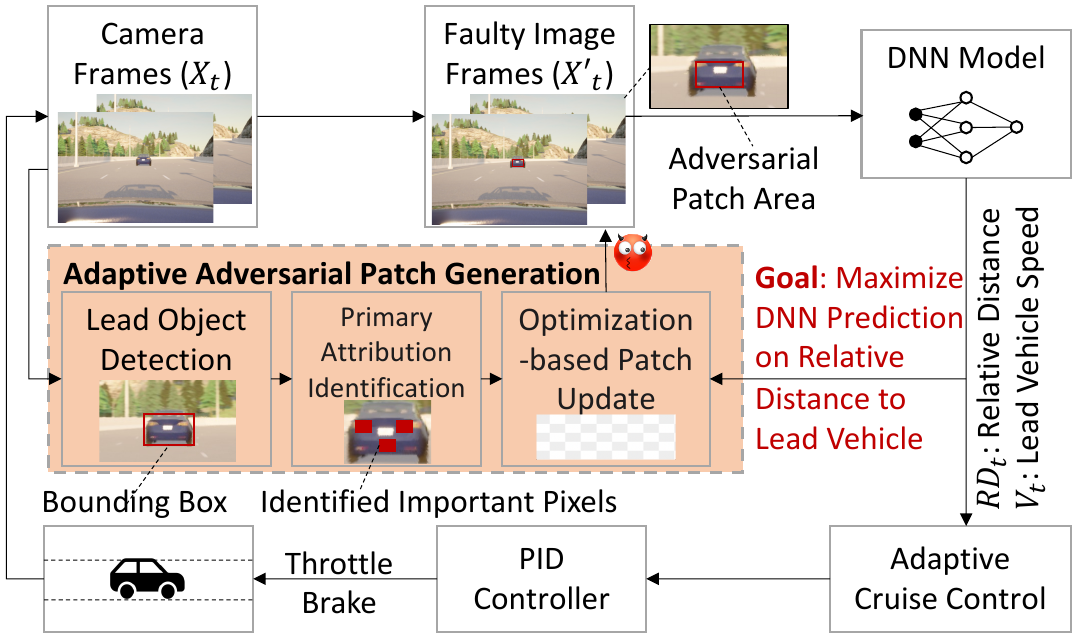}
    \vspace{-1em}
    \caption{Optimization-based adversarial patch generation.}
    \label{fig:opt}
    \vspace{-1em}
\end{figure}

The critical system contexts identified in the previous section are based on the high-level unsafe actions (e.g., Acceleration) issued by the ACC controller. In order to find the specific attack values or DNN input perturbations that can cause such unsafe control actions, we present an optimization-based patch generation method as shown in Fig. \ref{fig:opt}.  

\textbf{Runtime Optimization-based Adversarial Patch Generation.}
\label{subsec:objectivefunction}
To address challenge \textbf{C2},
we formulate the attack as the following runtime optimization problem:

\begin{align}
    \text{min }& \sum_{d\in {RD}_t}{-\triangledown g(d,\theta)} + \lambda ||\Delta_t ||_p \label{eq:obj}\\
    \textit{s.t. }
    & Patch_t = \Delta_t*M_t \label{eq:patch-matrix}\\
    &Patch_t \in [\mu-\sigma,\mu+\sigma] \label{eq:pvalue}\\
    & Area(Patch_t) \subset BBox(LV)_t \label{eq:plocation}\\
    &X^{adv}_t = X_t+Patch_{t} \label{eq:xupdate}\\
    & [{RD},RS]_t=LVD_{\theta}(X^{adv}_{t-1}) \label{eq:pvlead}\\
    & u_t = ACC(s_t,[{RD},{RS}]_t)\\
    & s_{t+1} = CarModel(s_t,u_t)
\end{align}
where Eq. \ref{eq:obj} is an objective function that aims at accelerating the Ego vehicle as soon as possible to cause a forward collision.
Directly decreasing the probability of a lead vehicle or its bounding box (BBox) cannot change the ACC system behavior.
We instead design an objective function that increases {$RD_t$} as much as possible while keeping the perturbation value of the adversarial patch imperceptible to human eyes. 

In Eq. \ref{eq:obj}, $g(d)$ is an approximate polynomial function of $d$ that fits the trend of the trajectory of the relative distance ${RD}_t$, predicted by the DNN model with weight parameters $\theta$. 
"-" is a negative sign that converts our goal of maximizing the relative distance to minimizing the proposed objective function. 
For example, when the gradient of the relative distance trajectory, $g(d)$, is negative, minimizing "$-\triangledown g(d)$" will slow down the decrease of $g(d)$ and is equivalent to maximizing the relative distance.
We adopt the gradient of $g(d)$ in the objective function instead of using ${RD}_t$ itself in order to avoid sharp changes in the predicted relative distance value, which might be easily detected by some anomaly detection mechanisms. 
{We assume the attacker has access to the DNN predictions (e.g., $RD$) by monitoring the ADAS communication network (e.g., ROS) or by running a replicated DNN model on a remote server or wireless communication device (see Table \ref{tab:threatmodel}).}


In Eq. \ref{eq:xupdate}, the perturbation is added to the original image input $X_t \in \mathbb{R}^{H\times W \times C}$ in the form of an adversarial patch $Patch_t \in \mathbb{R}^{H\times W \times C}$, 
represented as a matrix of pixels with height \textit{H}, width \textit{W}, and \textit{C} color channels. 
$\lambda$ is the weight parameter of the p-norm regularization term, designed to minimize the perturbation value of the patch for stealthiness. We limit the perturbation value within the Kalman filter noise parameters ($\mu$,$\sigma$) (Eq. \ref{eq:pvalue}), which ensures the perturbation is not corrected by the sensor fusion. We also constrain the adversarial patch inside the BBox of the LV (Eq. \ref{eq:plocation}) to enhance attack effectiveness, minimize the perturbation area for stealthiness, and reduce computational cost.

\begin{figure*}[t!]
\begin{minipage}[t]{0.8\textwidth} 
    \includegraphics[width=\textwidth]{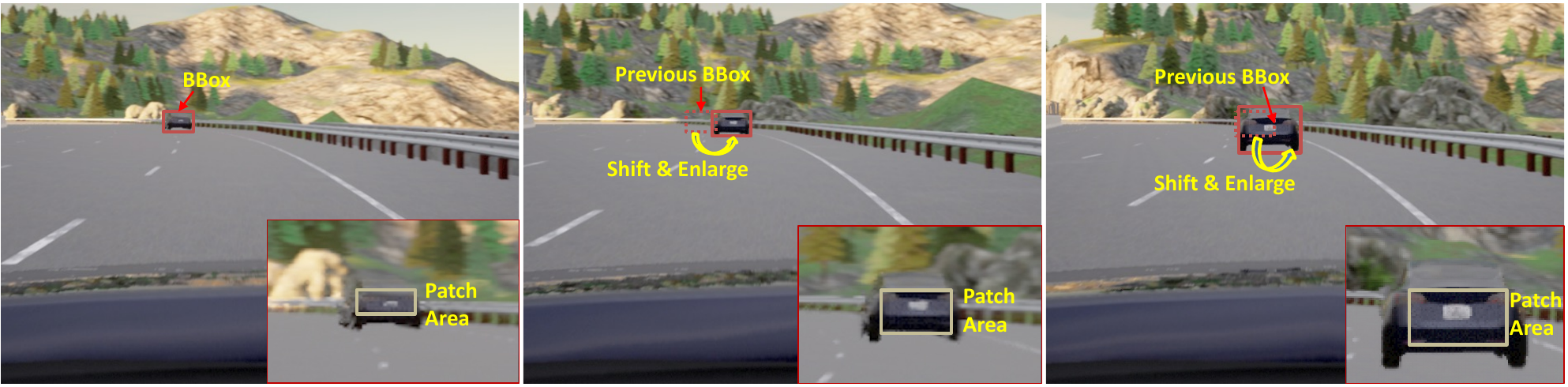}
    \vspace{-2em}
    \caption{Examples of the shift and adjustment process in the patch generation. 
    Inset figures are the zoomed-in views of the front vehicle with an adversarial patch added around the license plate area.} 
    \vspace{-0.5em}
    \label{fig:adjustpatch}
\end{minipage}
\hfill
\begin{minipage}[t]{0.17\textwidth}
\includegraphics[width=0.7\textwidth]{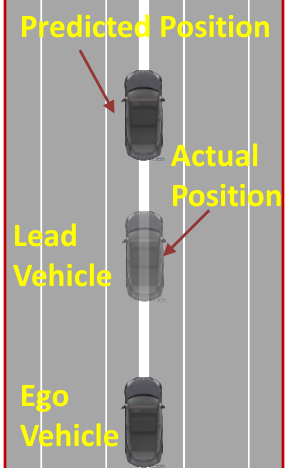}
\vspace{-1em}
\caption{ACC under attack.}
\vspace{-2em}
\label{fig:adsview}
\end{minipage}
\vspace{-0.5em}

\end{figure*}

\textbf{Primary Attribution Detection and Patch Update.}
\label{sec:patchadjust}
As mentioned in Section \ref{sec:challenges}, a major challenge (\textbf{C3}) in the design of runtime attacks is the changes in the size and location of the LV in the perceived image frames. To address this challenge, the attacker needs to update the generated patch dynamically according to the approximate DNN outputs. 
In this paper, we adopt an object detection method \cite{redmon2016you} to detect and track the real-time location and size of the LV and concentrate the attack perturbation within the detected BBox of the LV. In production ACC with object detection features \cite{Apollo}, given proper access, the attacker can skip this step and use the stock prediction results. 

After getting the BBox, we utilize a primary attribution algorithm \cite{sundararajan2017axiomatic} to quantify the relationship between input features and output predictions. Through this exploration, we try to identify the important pixels within the BBox of the LV that contribute the most to the predictions of {$RD_t$}. The input pixels with high weights identified by the attribution algorithm are marked by unit value in the mask matrix $M_t$, and the remaining pixels are assigned zero values. This mask matrix is then multiplied by the perturbation $\Delta_t$ to generate the adversarial $Patch_t$ (in Eq. \ref{eq:patch-matrix}). This step is useful as it can filter out non-important pixels in the inputs to reduce the number of perturbed pixels to improve the effectiveness of optimization-based attacks and reduce computation costs {of} runtime attacks.

Finally, we develop a new initialization algorithm to shift the patch position and adjust its size when the detected BBox changes (Eq. \ref{eq:patchshift}-\ref{eq:patchMask}). 
We shift the attack vector toward the new position of the detected BBox of the LV with a magnitude of ($x_t-x_{t-1}$, $y_t-y_{t-1}$), where ($x_{t-1}$, $y_{t-1}$) and ($x_t$, $y_t$) are the centers of BBox at previous and current control cycles, respectively (Eq. \ref{eq:patchshift}). We then expand the adversarial patch attack vector ($Patch$) to the dimension that matches the size of the newly detected BBox of the LV. Instead of reinitializing the whole attack vector matrix with random or zero values, which will reset the whole optimization process, we keep the previous patch values and intermediate variables and only initialize newly expended units (Eq. \ref{eq:patchInit}-\ref{eq:patchMask}). Fig. \ref{fig:adjustpatch} shows an example.
%
\begin{align}
  &\Scale[0.95]{Pos(Patch_t) = Pos(Patch_{t-1}) + (x_t-x_{t-1}, y_t-y_{t-1})} \label{eq:patchshift}\\
    &\Scale[0.95]{Init(\Delta_t) = [0]*size(BBox(LV)_t) +  \begin{bmatrix}
\Delta_{t-1} & 0 \\
 0& 0 \\
\end{bmatrix}} \label{eq:patchInit} \\
     &\Scale[0.95]{Init(Patch_t) = Init(\Delta_t)*M}  \label{eq:patchMask}
\end{align}

{This algorithm maintains a continuous optimization process across two consecutive perception cycles, which is critical in satisfying the real-time constraints. }Fig. \ref{fig:adsview} shows a visualization of how the adversarial patch affects the DNN predictions.



\section{Safety Intervention Simulation}
\label{sec:platform}
\begin{figure}[t]
    \centering
    \includegraphics[width=\columnwidth]{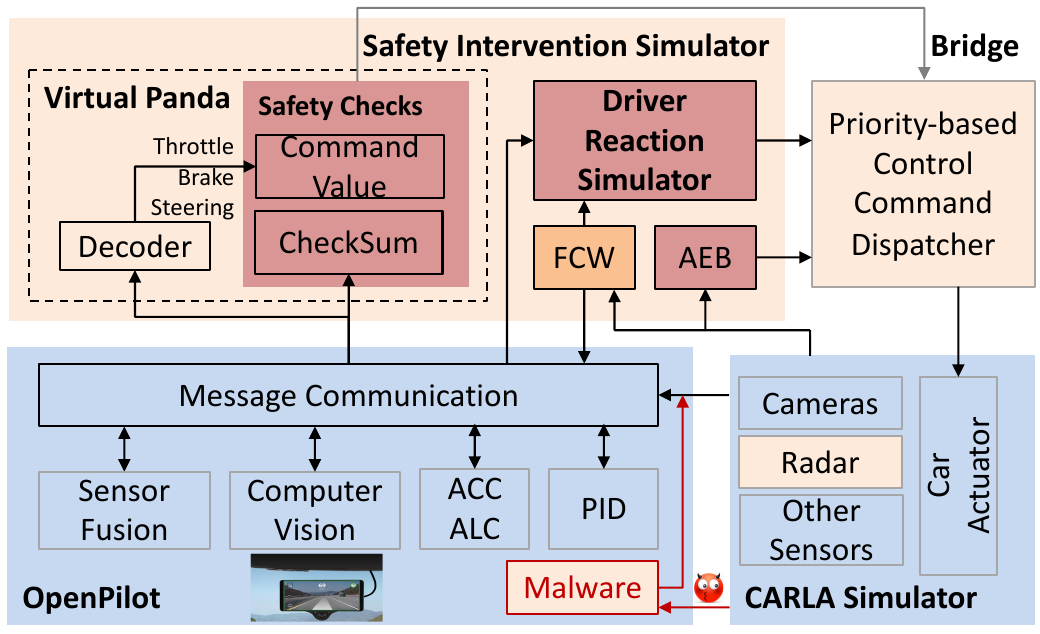}
    \vspace{-2em}
    \caption{Simulation platform. 
    [Code Available: https://github.com/gitguige/openpilot0.8.9]}
    \label{fig:platform}
    \vspace{-2em}
\end{figure}

To evaluate the safety of DNN-based ACC systems under attacks, we enhance the default OpenPilot and CARLA simulation platform (see Section \ref{sec:background:platform}) to be more representative of real-world ADAS, by developing a safety intervention simulator and mechanisms for priority-based dispatching of control commands to CARLA and fusion of camera and radar data (see Appendix \ref{appendix:fusion}). 
An overview of the simulation platform is shown in Fig. \ref{fig:platform} (with the orange parts representing our new implementations) and presented next.

To fill the gap in considering safety interventions and address the challenge of ensuring the combination of human driver and vehicle safe (see Section \ref{sec:bkground:safety}),
we implement and integrate three levels of safety interventions in the OpenPilot software (see Fig. \ref{fig:platform}), including ADAS safety features (e.g., AEB and FCW), basic car safety constraint checking on control commands, and driver interventions.

\textbf{AEBS (FCW and AEB) Simulator.}
\label{sec:AEB}
To design and test the AEBS mechanisms in simulation, we thoroughly review the regulations and requirements concerning AEBS~\cite{schram_implementation_nodate} \cite{http://data.europa.eu/eli/reg/2020/1597/oj} \cite{noauthor_grva-12-50r1epdf_nodate} and adhere to UN Regulation No. 152 \cite{http://data.europa.eu/eli/reg/2020/1597/oj}.
We adopt and implement a time-to-collision (TTC) based phase-controlled AEBS \cite{alsuwian_autonomous_2022} in our platform. 


The AEBS processes inputs from LVD outputs (after sensor fusion), including relative distance ($RD$) and relative speed ($RS$), and current speed $V_{Ego}$ (see Fig. \ref{fig:AdasOverview}). 
The average driver reaction time ($T_{react}$) is set to the commonly used constant value of 2.5s \cite{zhou2022strategic, sato2021dirty}. Time thresholds, namely $ttc$ (time to collision), $t_{fcw}$ (FCW time), $t_{pb1}$ (1st phase partial brake time), $t_{pb2}$ (second phase partial brake time), and $t_{fb}$ (full brake time), are then calculated as follows:
\begin{align}
\label{eq:ttc_logic}
      ttc &= RD / RS; \\
      t_{fcw} &= T_{react} + V_{Ego} / 4.5 \\
      t_{pb1} &= V_{Ego} / 2.8; 
      t_{pb2} = V_{Ego} / 5.8; 
      t_{fb} = V_{Ego} / 9.8 
\end{align}

\begin{figure}[t]
    \centering
    \begin{minipage}{0.345\columnwidth}
    \includegraphics[width=\columnwidth]{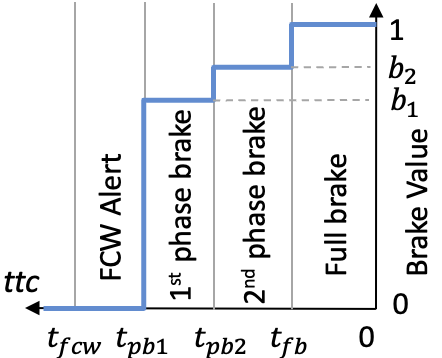}
    \vspace{-2em}
    \caption{AEBS.}
    \label{fig:aebstimeline}
    \end{minipage}
    \hfill
    \begin{minipage}{0.635\columnwidth}
    \includegraphics[width=\columnwidth]{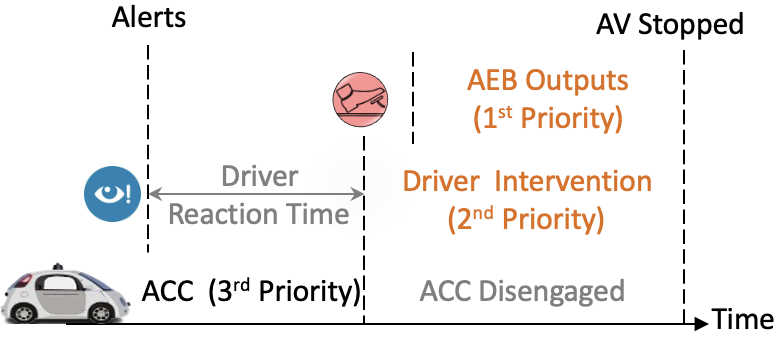}
    \vspace{-2em}
    \caption{Control command dispatcher.}
    \label{fig:comdispatcher}
     \end{minipage}
     \vspace{-2.3em}
\end{figure}

    
As shown int Fig. \ref{fig:aebstimeline}, when $ttc$ falls below $t_{fcw}$, $t_{pb1}$, $t_{pb1}$, and $t_{pb1}$, a corresponding action (warning or brake with 90\%, 95\%, 100\% force) is executed. 
Applying the brake blocks other ADAS controls. 
See Appendix \ref{appendix:aeb} for more details on AEBS design and testing.

In reality, when OpenPilot is installed on a car, some car models lose the AEBS functionality \cite{Carmodels-OpenPilot}, while others retain it. Also, AEBS might rely on a separate ADAS camera \cite{adas-camera-can}, distinct from the OpenPilot camera, and a potential compromise of the AEBS camera data is possible. Thus, we consider three scenarios for AEBS interventions: (1) AEBS is enabled, and AEBS camera data is uncompromised; (2) AEBS is enabled, but AEBS camera data is compromised; and (3) AEBS is disabled (see Section \ref{sec:evadingsafety} and Table \ref{tab:reswithsafety}).






\textbf{Safety Constraint Checker.}
The OpenPilot safety mechanisms are implemented in its control software and the Panda CAN interface. Panda is a universal OBD adapter developed by Comma.ai~\cite{Panda} that provides access to almost all car sensors through the CAN bus and also enforces some safety constraints over output commands. However, when integrated with the CARLA driving simulator, OpenPilot does not utilize Panda software {or hardware; thus} Panda safety checks are inactive.

To be as realistic as the actual OpenPilot on the road, we add a \textit{virtual Panda} module in our simulation that copies the exact logic of Panda software \cite{Panda}. 
Specifically, as shown in Fig. \ref{fig:platform}, the virtual Panda decodes the CAN packages sent by the ACC and checks whether their checksum is correct and the control command values are within predefined thresholds \cite{Panda}. For example, to ensure safety, the maximum acceleration and deceleration of the vehicle shall be limited between 2$m/s^2$ and -3.5$m/s^2$ respectively \cite{openpilot_safety}. Only the commands that pass the Panda safety checks are sent to the simulated vehicle actuators.
In CARLA simulator, the final control commands are truncated within the range of [0,1].

\begin{table}[t]
\caption{Driver simulator: activation conditions and reactions.}
\vspace{-1em}
\label{tab:driverReaction}
\resizebox{\columnwidth}{!}{%
\begin{tabular}{@{}lll@{}}
\toprule
\textbf{Activation Condition} & \textbf{Driver Reaction} & \textbf{Reaction Time} \\ \midrule
Alerts (e.g., FCW) &
  \multirow{5}{*}{\begin{tabular}[c]{@{}l@{}}Emergency Brake \& Zero Throttle \\ No changes in the steering angle\end{tabular}} &
  \multirow{5}{*}{2.5 seconds} \\
Unexpected Acceleration &                 &               \\
Unsafe Cruise Speed     &                 &               \\ 
{Unsafe Following Distance} &&  \\
Obvious Camera Perturbation && \\\midrule
\multirow{2}{*}{Hard Braking} &
  \multirow{2}{*}{\begin{tabular}[c]{@{}l@{}}Stop brake and output regular throttle\\ No changes in the steering angle\end{tabular}} &
  \multirow{2}{*}{2.5 seconds} \\
                        &                 &               \\ \bottomrule
\end{tabular}%
}
\vspace{-1.7em}
\end{table}

\textbf{Driver Reaction Simulator.}
\label{sec:driver}
To assess driver interventions, we develop a driver reaction simulator. 
The simulated driver is notified when any safety alerts are raised by the ADAS (e.g., FCW) or when the driver observes any abnormalities in the vehicle's status or camera user interface (UI). These ACC abnormalities include hard braking, unexpected acceleration, the vehicle's speed exceeding the cruising speed by more than 10\%, {unsafe following distance with the lead vehicle (e.g., less than a vehicle length)}, or the mean perturbation value in the UI surpassing a noticeable threshold, with a default value of 15\% representing an alert driver ($Patch.mean()>0.15$).
We assume a very alert driver who can notice any anomalies that occur within a single control cycle (10ms).
A predefined emergency response will be issued by the driver accordingly (see Table \ref{tab:driverReaction}), taking effect 2.5 seconds later (average driver reaction time).
To mimic the human driver's braking behavior, we adopt a braking curve function from previous research \cite{zhou2022strategic}. 

\textbf{Priority-based Control Command Dispatcher.} With multiple safety mechanisms in place, there might be conflicts among the control commands issued by the OpenPilot ACC controller and those generated by the safety interventions. To resolve such conflicts,
we design a command dispatcher to transmit output control commands to the CARLA actuators from various sources (e.g., ACC, AEB, simulated driver) based on their priorities, with high-priority commands overwriting low-priority ones (see Fig. \ref{fig:platform} and Fig. \ref{fig:comdispatcher}). The simulated driver's actions have a higher priority than regular ACC outputs, and control actions from the AEB have the highest priority. 
The driver's actions will be executed 2.5s (average driver reaction time) after safety alerts (e.g., FCW) or noticing other ACC malfunctions (see Table \ref{tab:driverReaction}). ACC commands will be blocked or disengaged when AEB or driver interventions are triggered.

\section{Evaluation in Simulated Environment}
\label{sec:evaluation}
This section presents the evaluation of our proposed context-aware attack strategy (referred to as \textbf{CA-Opt}) using the enhanced simulation platform presented in Section \ref{sec:platform}.


\subsection{Methodology}
\label{sec:expsetup}


We study the following research questions by comparing the effectiveness of CA-Opt attack to several baseline attack methods in causing safety hazards (Section \ref{sec:evaluation:effectiveness}) and evading different safety interventions (Section \ref{sec:reswithsafety}): 

\textbf{RQ1:} Does strategic selection of attack times and values increase the chance of hazards (forward collisions)?

\textbf{RQ2:} Does stealthiness design help maintain the attack effectiveness in the presence of safety interventions?

{\textbf{RQ3:} Does a perception input attack achieve better performance than direct {perception and control output attacks}?}

\textbf{Baselines.} 
We design three
baseline attack strategies to answer these questions (see Table \ref{tab:baselines}). 
\begin{table}[b]
\vspace{-1em}
\caption{Overview of proposed and baseline attack strategies. }
\vspace{-1em}
\label{tab:baselines}
\resizebox{\columnwidth}{!}{%
    
\begin{tabular}{@{}lllll@{}}
\toprule
\textbf{Attack} & \textbf{Start Time} & \textbf{Duration} & \textbf{Attack Value} & \textbf{$\#$Sim.} \\ \midrule

CA-Random & Context-Aware & Context-Aware & Random & 1000 \\
CA-APGD & Context-Aware & Context-Aware & AutoPGD & 1000 \\
Random-Opt & Uniform {[}5,40{]}s & Uniform {[}0.5,2.5{]}s & Opt-based & 1000 \\ 
CA-Opt (Ours) & Context-Aware & Context-Aware & Opt-based & 1000 \\ 
\bottomrule
\end{tabular}%


}
\end{table}


%

To assess the effectiveness of our optimization-based adversarial patch method in strategically selecting the attack values, we compare it to two baselines: \textbf{CA-Random}, which introduces random perturbations to perception inputs, and \textbf{CA-APGD}, which uses a state-of-the-art gradient-based method, Auto-PGD \cite{croce2020reliable}, to determine the perturbation values. {Since the original Auto-PGD is designed for misclassification, which does not work for attacking ACC, we change the goal function {(to $\triangledown g(d)$, see Eq. \ref{eq:obj})} to maximize relative distance prediction. We also limit the iteration number to 5, the maximum number of control cycles (100Hz) within a perception cycle (20Hz)).}
Both baselines use the same context-aware method as the proposed CA-Opt attack strategy to choose the attack start time and duration. To evaluate our method's efficiency in selecting the timing and duration of attacks, we design another baseline (\textbf{Random-Opt}) that selects a random start time uniformly distributed within [5, 40] seconds and a random attack duration uniformly distributed within [0.5, 2.5] seconds. To concentrate only on the effect of different start times and durations, Random-Opt shares the same adversarial patch generation method as CA-Opt.
For a fair comparison, we confine perturbations to the detected bounding box (BBox) of the LV and update BBox size and position with our proposed patch updating algorithm (Section \ref{sec:patchadjust}).
%



\textbf{Driving Scenarios.}
We model a 2016 Honda Civic, both with and without basic safety features, navigating curvy and straight sections of a highway using the "Town04\_opt" map under clear weather and dry road conditions in CARLA. We simulate four high-risk driving scenarios designed based on the NHTSA's pre-collision scenario topology report \cite{najm_pre-crash_2007}. The Ego vehicle, traveling at 60 mph and positioned 75 meters away, encounters a LV exhibiting various behaviors: 
(SC1) LV cruises at the speed of 35 mph;
(SC2) LV cruises at the speed of 50 mph;
(SC3) LV slows down from 50 mph to 35 mph; and
(SC4) LV accelerates from 35 mph to 50 mph.

Our experiments are done on Ubuntu 20.04 LTS, with OpenPilot v0.8.9 and CARLA v9.11. A single simulation of OpenPilot contains 5,000 time steps, and each step lasts about 10 ms, which equals 50 seconds in total. {However, if an attack leads to a collision, the simulation ends earlier.}
For Random-Opt, we randomly select ten start times and five durations for each test scenario and repeat them five times, which results in 1,000 simulations. The same total number of simulations is done for CA-Opt and other baselines.


\subsection{Attack Success Rate in Causing Hazards}

\label{sec:evaluation:effectiveness}



To assess the CA-Opt attack's effectiveness in inducing safety hazards (\textbf{RQ1}), we conduct experiments in the closed-loop simulation platform without enabling the safety interventions.
This setting is similar to previous works \cite{sato2021dirty,guo2023adversarial}.

We consider an attack successful if a collision event is observed (Ego vehicle collides with LV) or relative distance between LV and Ego vehicle is no larger than 0m. {The success rate is reported across 1,000 simulations, if not specified otherwise.}
\label{sec:Results:attackValue}
\begin{figure}[t]
    \centering
	\scriptsize \centering
    \includegraphics[width=\columnwidth]{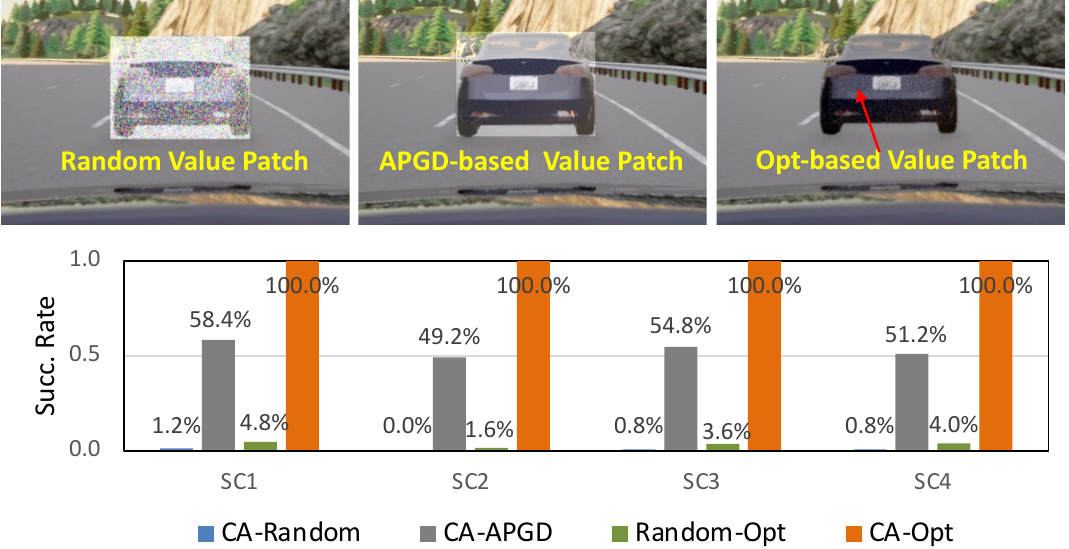}
    \vspace{-2em}
    \caption{Top: Adversarial patch examples generated using proposed method vs. random and APGD-based methods. Bottom: Success rate of  CA-Opt and baseline attacks in absence of safety interventions.} 
    \vspace{-2em}
    \label{fig:resOpt}
    \vspace{-1em}
\end{figure}

Fig. \ref{fig:resOpt} displays the success rates for each attack. The CA-Random attack leads to hazards in less than 1.2\% of scenarios, with an overall success rate of 0.7\%. This low rate suggests that randomly generated adversarial patches minimally impact the DNN model predictions and rarely cause hazards. Increasing the random perturbation values of the adversarial patch also does not significantly enhance the success rate. This is because OpenPilot's DNN model is primarily trained to detect the front objects but not to identify their specific class, thus showing greater resilience to adversarial attacks. Additionally, patches with larger perturbation values are more visible to the driver than those from the optimization-based method (Fig. \ref{fig:resOpt}-Top), potentially alerting the driver to prevent hazards.

In contrast, the proposed CA-Opt attack achieves a 100\% success rate in all testing scenarios, surpassing CA-Random by 142.9 times. 
{Although CA-APGD uses a similar goal function as CA-Opt, it does not cause hazards in 46.6\% of simulations, mainly due to limiting its number of iterations to satisfy real-time constraints. On the other hand, by employing a dynamic patch updating algorithm, the CA-Opt attack ensures the continuity of the optimization process across the perception cycles and enhances the attack's effectiveness.}

We also observe that the Random-Opt baseline only achieves an average success rate of 3.5\%, 28.6 times lower than the CA-Opt, as it wastes resources by injecting perturbations at non-critical system states. This highlights the importance of strategic timing of attacks and the insufficiency of optimization-based methods alone in causing hazards.

\noindent\vsepfbox{
    \parbox{0.95\linewidth}{
    \textbf{Observation 1: }CA-Opt is more efficient than baselines in identifying the most critical times and optimal DNN perturbation values for attacking the ACC systems at runtime and {overcoming real-time constraints (C3)}.
    }}

\subsection{Attack Stealthiness with Safety Interventions}
\label{sec:reswithsafety}

This section studies the impact of the safety interventions and the stealthiness design on attack efficiency (\textbf{RQ2}).

\subsubsection{Stealthiness in Perception Input}
\label{sec:evaluation:stealthuness}
To evade detection by safety mechanisms and human driver, the adversarial patch should stay as stealthy as possible.
Basically, the smaller the value of the pixels' perturbations, the stealthier the attack will be. Therefore, we tested our attack method with three different $\lambda$ values in Eq. \ref{eq:obj}. 
We use two sets of metrics to evaluate the stealthiness of the patch, including (i) the degree of the pixel perturbation measured using $L_2$ and $L_\infty$ distance \cite{norm} and (ii) the similarity between the original camera image and the perturbed image, calculated using RMSE and universal image quality index (UIQ) \cite{wang2002universal}. 

Table \ref{tab:resStealth} presents the results averaged over all the test scenarios and simulations. The CA-Opt attack achieves at least a 99.2\% success rate under all three stealthiness levels and keeps the perturbation degree less than 0.015 ($L_\infty$) and 0.184 ($L_2$). The perturbed image with the adversarial patch has a similarity of UIQ = 0.993 (1 means identical) to the original image. We choose the $\lambda$ value to be $10^{-3}$ in our evaluations because of its stealthiness and high attack effectiveness. Examples of the generated adversarial patches (with $\lambda=10^{-3}$) are presented in Fig. \ref{fig:adjustpatch} (see the zoomed-in area) and Fig. \ref{fig:resOpt}, which are almost invisible to human eyes.

\begin{table}[b]
\centering
\small
\vspace{-2em}
\caption{Attack success rate with different patch stealthiness levels. 
}
\vspace{-1.5em}
\label{tab:resStealth}
\resizebox{\columnwidth}{!}
{%
\begin{threeparttable}
    
\begin{tabular}{@{}llllll@{}}
\toprule
\multirow{2}{*}{\begin{tabular}[c]{@{}l@{}}\textbf{Stealthiness}\\ \textbf{Level} $\lambda$\end{tabular}} & \multirow{2}{*}{\begin{tabular}[c]{@{}l@{}}\textbf{Succ.}\\ \textbf{Rate}\end{tabular}}& \multicolumn{2}{c}{\textbf{Perturbation Pixel}}  & \multicolumn{2}{c}{\textbf{Image Similarity}} \\ \cmidrule(l){3-4} \cmidrule(l){5-6} 
 &  &  $L_2$ & $L_\infty$ & RMSE($\times10^{-5}$) & UIQ  \\ \midrule
$10^{-2}$ & 99.2\%  &0.086  & 0.015  & 1.061 & 0.993 \\
$10^{-3}$ & 100\%  &0.128  & 0.015  &  1.168 & 0.993  \\
$10^{-4}$ & 100\%  &  0.184 & 0.015 & 1.319 &  0.993  \\ \bottomrule
\end{tabular}%

\begin{tablenotes}
\footnotesize
    \item[*] $L_2$ and $L_\infty$ distances are the normalized perturbation values of the attack vector matrix in the range of [0,1]. 
    Image similarity is evaluated by comparing the RMSE and UIQ between the original image and the perturbed image with the patch. Smaller RMSE and larger UIQ mean higher similarity.
\end{tablenotes}

\end{threeparttable}

}

\end{table}

To further evaluate the stealthiness of our attack design, we also conduct a user study with 30 participants. Results show that adversarial patches at $\lambda=10^{-2}$ and $\lambda=10^{-3}$ are almost imperceptible to human drivers, and the patches generated by CA-Opt attacks are less noticeable than those generated by baseline perception attacks (CA-Random and CA-APGD) (refer to Appendix \ref{sec:appendix:userstudy} for more details).

\subsubsection{Evading Safety Interventions} 
\label{sec:evadingsafety}
For a more realistic evaluation of the effectiveness of different attack strategies, we rerun our experiments with different safety interventions (introduced in Section \ref{sec:platform}). 
A calibration of the safety features is performed before the experiments to ensure the interventions are triggered correctly without any false positives.
We test each attack method with different AEBS configurations: (i) AEB/FCW depends on an independent camera that is not compromised, (ii) AEB/FCW utilizes compromised camera inputs similar to the ACC (simulating stock ACC and AEBS that share a camera or independent ACC and AEBS cameras that are both compromised), or (iii) AEB/FCW is disabled.
Driver intervention and ACC safety constraint checking (OpenPilot Panda checks) are considered for all three settings. Here, we do not test the Random-Opt attack due to its similarity to CA-Opt but with worse performance. 
We assess the efficacy of each attack method using metrics such as the attack success rate, safety intervention activation rate (indicating the percentage of simulations triggering safety interventions), and hazard prevention rate (the percentage of simulations where hazards occur without safety interventions). 
Table \ref{tab:reswithsafety} shows the experimental results of each attack method with different safety intervention configurations. 
We observe that, regardless of the interventions, the CA-Random and CA-APGD attacks fail to cause any hazards due to their low baseline success rates (see Fig. \ref{fig:resOpt}) and their noticeable perturbations that trigger the driver interventions in 23.8-27.4\% and 100\% of scenarios. 
These findings highlight the effectiveness of human drivers in preventing accidents and keeping autonomous driving safe.  

\begin{figure}[t!]
    \centering
    \includegraphics[width=0.7\columnwidth]{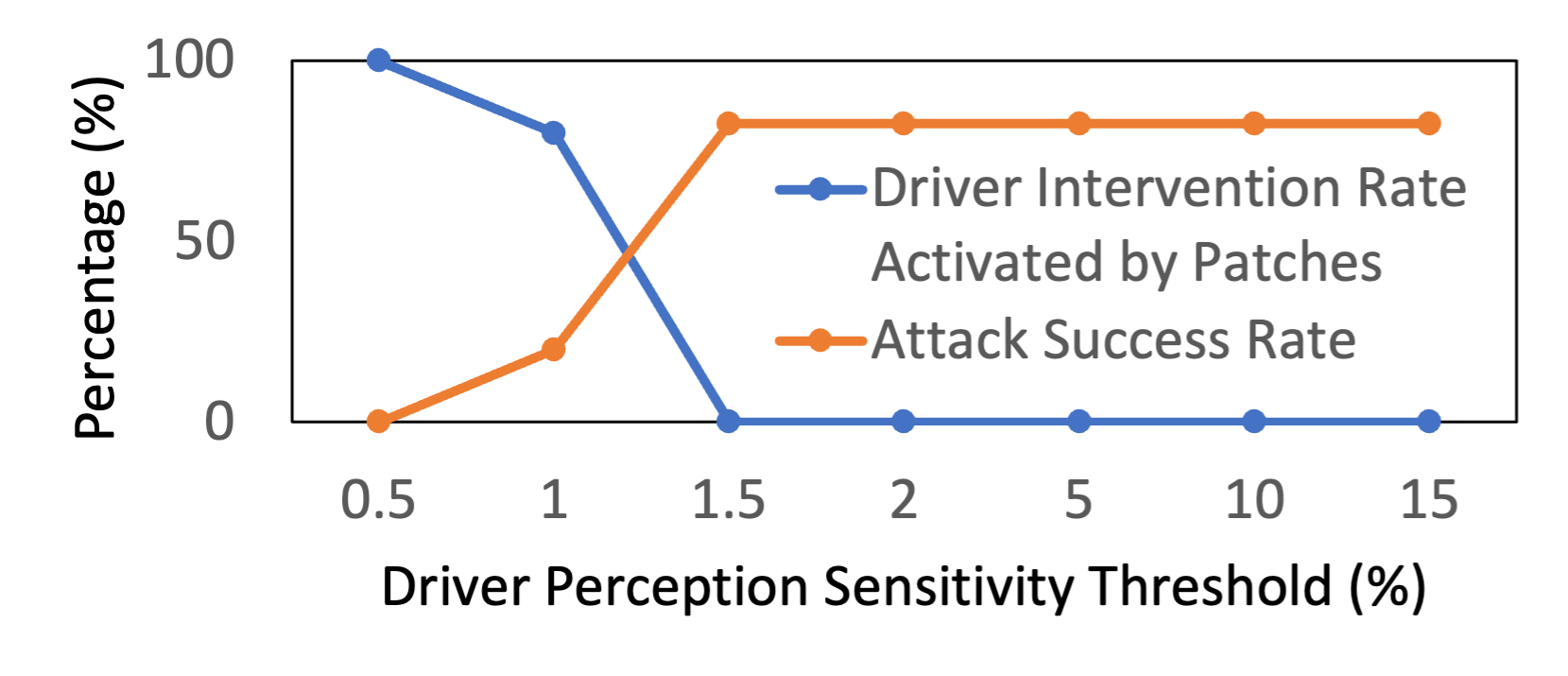}
    \vspace{-1.5em}
    \caption{Evaluation with different driver sensitivity thresholds.}
    \label{fig:driverthresholds}
    \vspace{-2em}
\end{figure}

In contrast, with AEBS disabled, the CA-Opt attack resulted in an average attack success rate of 82.6\%. 
%
We also conduct experiments that simulate higher driver sensitivity levels by decreasing the mean perturbation value threshold for activating driver intervention from the default value of 15\% (see Section \ref{sec:driver}) to 10\%, 5\%, 2\%, 1.5\%, 1\%, and 0.5\%. As shown in Fig. \ref{fig:driverthresholds}, when perturbation thresholds are set to 0.5\% and 1\% (representing highly sensitive driver), the adversarial patch triggers driver interventions in all and 79.6\% of the simulations, leading to attack success rates of 0\% and 20.4\%, respectively. However, with thresholds higher than 1.5\%, our attack maintains an 82.6\% success rate. 
This finding underscores the robustness of the generated adversarial patch in evading driver detection across a range of driver sensitivities. 

\begin{table}[t!]
\footnotesize
\caption{Performance of attacks with all the safety features and different AEBS settings.}
\vspace{-1em}
\centering
\label{tab:reswithsafety}
\resizebox{\columnwidth}{!}
{%
\begin{tabular}{@{}lllll@{}}
\toprule
\begin{tabular}[c]{@{}l@{}}\textbf{Safety} \\ \textbf{Interventions}\end{tabular} &
\begin{tabular}[c]{@{}l@{}}\textbf{Attack} \\ \textbf{Method}\end{tabular} &
  
  \begin{tabular}[c]{@{}l@{}}\textbf{Intervention} \\ \textbf{Activation Rate}\end{tabular} &
  \begin{tabular}[c]{@{}l@{}}\textbf{Succ.} \\ \textbf{Rate}\end{tabular} &
  \begin{tabular}[c]{@{}l@{}}\textbf{Hazard} \\ \textbf{Prevention Rate}\end{tabular}  \\ \midrule
\multirow{3}{*}{\begin{tabular}[c]{@{}l@{}}All \& \\ AEBS Not Compromised \\(Independent Camera)\end{tabular}}& CA-Random  & 27.4\% & 0 & 100\% (7/7) \\
                        & CA-APGD    & 100\%  & 0 & 100\% (534/534) \\
                        & CA-Opt     &100\%  & \textbf{48.7\%} &\textbf{51.3\%} (513/1000)   \\ 
                        
                        \midrule

\multirow{3}{*}{\begin{tabular}[c]{@{}l@{}}All \& \\ AEBS Disabled/  \\ Compromised (Shared Camera)\end{tabular}}& CA-Random     & 23.8\%/ 24.3\% & 0 & 100\% (7/7)   \\
                        & CA-APGD    & 100\%  & 0 &  100\% (534/534)  \\
                        & CA-Opt     & {100\%}  & \textbf{82.6\% }&  \textbf{17.4\%} (174/1000)  \\ 
                        
\bottomrule
\end{tabular}%
}
\vspace{-2em}
\end{table}

When AEBS is enabled and uses the same compromised camera inputs as ACC, we observe a similar high success rate (82.6\%) for the CA-Opt attacks that affect both the ACC and AEBS functionalities. 
However, the CA-Opt attack encounters challenges when the AEBS relies on uncompromised camera data from an independent camera. In this scenario, the attack triggers AEBS interventions in all simulations. But it still maintains a success rate of 48.7\% through a gradual (stealthy) change in the vehicle state (see Fig. \ref{fig:GasAccSpeed}) that delays AEBS activation and leaves insufficient time for hazard prevention. 

\vspace{0.5em}
\noindent\vsepfbox{
    \parbox{0.95\linewidth}{
        \textbf{Observation 2}: Our simulated safety interventions are effective in preventing accidents, and as required for L2 AVs, the human driver should always be in the loop and actively monitor ADAS to ensure safety. 
    }
}

\noindent\vsepfbox{
    \parbox{0.95\linewidth}{
        \textbf{Observation 3}: CA-Opt attack is more effective than baselines in keeping perturbations stealthy and causing hazards without being mitigated by safety interventions.
    }
}



\subsection{Comparison to DNN Output and Control Output Attacks}
\label{sec:perception_vs_output}

The stealthy perturbations on the perception input can get propagated through the DNN model and ACC logic and lead to changes in {the DNN output (\tikz[baseline=(char.base)]{\node[shape=circle,draw=Mahogany,fill=Mahogany,text=white,inner sep=1pt] (char) {2}} in Fig. \ref{fig:AdasOverview}) and} ACC control output \tikz[baseline=(char.base)]{\node[shape=circle,draw=Mahogany,fill=Mahogany,text=white,inner sep=1pt] (char) {3}}. Although the attacker's goal is to {maximize errors in DNN output and} cause sudden accelerations on the ACC output, large deviations in vehicle states may be detected by the human driver or existing safety and defense mechanisms. {To further evaluate the stealthiness of our proposed attack (CA-Opt), we compare deviations resulting from the attack to those caused by stealthy attacks directly on DNN and control outputs. Note such attacks are only possible under specific threat models (e.g., malware or wireless methods) in Table \ref{tab:threatmodel}. }



{\textbf{Control Output Attacks.}} We first examine deviations in the autonomous vehicle states and control outputs resulting from the attack compared to two baseline output attacks, called \textbf{MaxOut} and \textbf{StrategicOut}. These attacks directly modify ACC output control commands, by setting them to a maximum allowed acceleration value (MaxOut) or a strategic value (StrategicOut) based on a method from prior research \cite{zhou2022strategic}. But they use the same context-aware method as CA-Opt for selecting attack times and durations. 

\begin{figure}[t!]
    \centering
    \includegraphics[width=.9\columnwidth]{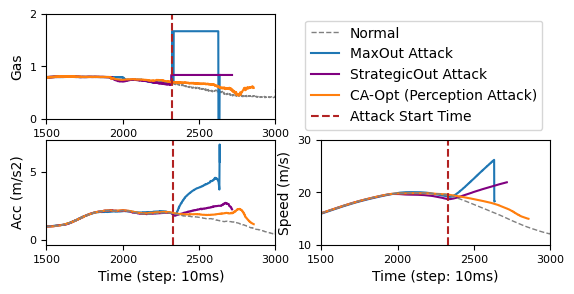}
    \vspace{-1.5em}
    \caption{Context-Aware perception attacks vs. output attacks.}
    \label{fig:GasAccSpeed}
    \vspace{-1.5em}
\end{figure}

Fig. \ref{fig:GasAccSpeed} illustrates an example scenario. The MaxOut attack leads to faster collisions, but also results in more noticeable changes in critical states such as gas, acceleration, and vehicle speed. These significant alterations are easily detectable by anomaly detection mechanisms or can be promptly noticed and addressed by human drivers.
In contrast, fixed perturbations injected by CA-Opt attack to DNN perception inputs may not propagate to cause any changes in ACC output or if they cause any changes, it will not be larger than the maximum possible acceleration caused by MaxOut attacks. These perturbations lead to gradual deviations of system states over a longer time period, thus achieving a high success rate (as shown in Fig. \ref{fig:resOpt}) while reducing the likelihood of detection. 
Although StrategicOut produces smaller deviations strategically to avoid safety alerts, changes in vehicle states (e.g., speed) are still more noticeable than the CA-Opt perception attack.

We also evaluate the success rate of StrategicOut attack under two different safety intervention configurations. 
We do not assess the MaxOut attack due to its high likelihood of being detected.
Table \ref{tab:reswithsafety2} shows that without AEBS, StrategicOut achieves a success rate of 81.9\% by generating attack values within safety limits and avoiding driver intervention. However, with AEBS active, using the same camera inputs as ACC, the success rate drops significantly to 20.3\%, due to AEBS interventions triggered in all simulations. 

\begin{table}[t!]
\caption{Performance of StrategicOut attack with all the safety features and different AEBS settings (AEBS with Shared Camera).}
\vspace{-1em}
\centering
\label{tab:reswithsafety2}
\resizebox{0.95\columnwidth}{!}
{%
\begin{tabular}{@{}llll@{}}
\toprule
\begin{tabular}[c]{@{}l@{}}\textbf{Safety} \\ \textbf{Interventions}\end{tabular} &
  
  \begin{tabular}[c]{@{}l@{}}\textbf{Attack} \\ \textbf{Method}\end{tabular} &
  \begin{tabular}[c]{@{}l@{}}\textbf{Succ.} \\ \textbf{Rate}\end{tabular} &
  \begin{tabular}[c]{@{}l@{}}\textbf{Hazard} \\ \textbf{Prevention Rate}\end{tabular}  \\ \midrule

All \& AEBS Activated      &  StrategicOut & 20.3\% & 79.7\% (797/1,000) \\

All \& AEBS Disabled     &  StrategicOut & {81.9\%} &  18.1\%(181/1,000)\\ \midrule

{All \& AEBS Activated}      & OptOut& {34.5\%} & {65.5 (655/1,000) } \\
\bottomrule
\end{tabular}%
}
\vspace{-1.2em}
\end{table}

{We further compared the CA-Opt attack with a stealthy control output attack that causes the exact deviations of the state variables as the proposed perception attack (referred to as \textbf{OptOut}). Specifically, we reran the simulations and injected the faults by setting the control output to the recorded output traces caused by the CA-Opt perception attack.
We observed that the OptOut attack achieved a higher success rate (34.5\%) than the StrategicOut attack. However, it did not change the DNN predictions or affect the AEBS function, thus triggering safety interventions more easily and earlier than the CA-Opt attack. 
In addition, CAN outputs are encrypted in some car models \cite{jha2020ml}, increasing the attack cost.}

\begin{figure}[t]
    \centering
    \includegraphics[width=0.65\columnwidth]{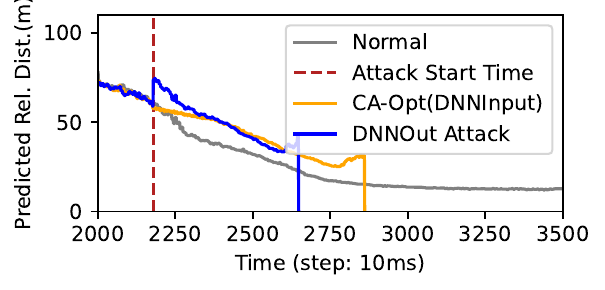}
    \vspace{-1.5em}
    \caption{CA-Opt perception attack vs. DNN output attack.}
    \vspace{-1em}
    \label{fig:dnnout}
\end{figure}

\textbf{DNN Output Attacks.} Similarly, we compare CA-Opt attack with a stealthy attack that directly compromises DNN output \tikz[baseline=(char.base)]{\node[shape=circle,draw=Mahogany,fill=Mahogany,text=white,inner sep=1pt] (char) {2}} (referred to as \textbf{DNNOut}) {by formulating an optimization problem to maximize the RD prediction within one standard deviation while ensuring the satisfaction of safety constraints on acceleration and speed \cite{zhou2022strategic}}. We calculate the acceleration and speed values corresponding to RD predictions by replicating the Openpilot MPC and PID algorithms.
This baseline uses the same context-aware method as CA-Opt for selecting the attack times and durations.
We observe that the DNNOut attack causes a more obvious change in the RD predictions {(see Fig. \ref{fig:dnnout})} compared to CA-Opt attack on DNN inputs \tikz[baseline=(char.base)]{\node[shape=circle,draw=Mahogany,fill=Mahogany,text=white,inner sep=1pt] (char) {1}}, which then results in similar obvious changes in the gas, speed, or acceleration as depicted in Fig. \ref{fig:GasAccSpeed}. 


\noindent\vsepfbox{
    \parbox{0.95\linewidth}{
    \textbf{Observation 4: }CA-Opt attack has advantage over direct DNN or control output attacks in minimizing vehicle state changes to evade detection by safety interventions, while maintaining high effectiveness in causing hazards.
    }    
}

\section{Evaluation in Real-World Settings}

In this section we aim to answer the following questions about the effectiveness of our attack in real-world settings. 


\textbf{RQ4:} Can our attack transfer well from simulation to real-world implementation?


\textbf{RQ5:} Can our attack evade detection or mitigation by the existing adversarial patch defense methods?

We also have evaluated the runtime overhead of our attack and its robustness to real-world factors, as described in Appendix \ref{sec:robustness}-\ref{sec:timeliness}.

\subsection{Performance on Actual Vehicles}
\label{sec:sim-to-real}
{We assess the feasibility of the CA-Opt attack on an actual vehicle (Lexus NX 2020) equipped with a production L2 ADAS, Comma 3, running OpenPilot software v0.8.9} by examining the attack impact on (i) the DNN perception module only and (ii) the end-to-end ACC. 


\textbf{Perception Module Evaluation.}
We evaluate the perception module in two scenarios of approaching an LV when (i) parked in a parking lot and (ii) driving on an actual road. In each scenario, the ACC on the Ego vehicle was tested with and without the adversarial patches injected to the camera frames.

\begin{figure}[t]
    \centering
    \begin{minipage}{0.3\columnwidth}
        \small
        \centering
        \includegraphics[width=\columnwidth]{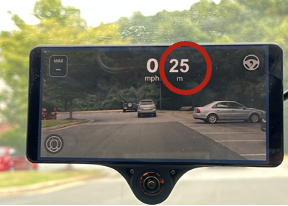}
        (1a)  
    \end{minipage}
    \hfil
    \begin{minipage}{0.3\columnwidth}
        \small
        \centering
        \includegraphics[width=\columnwidth]{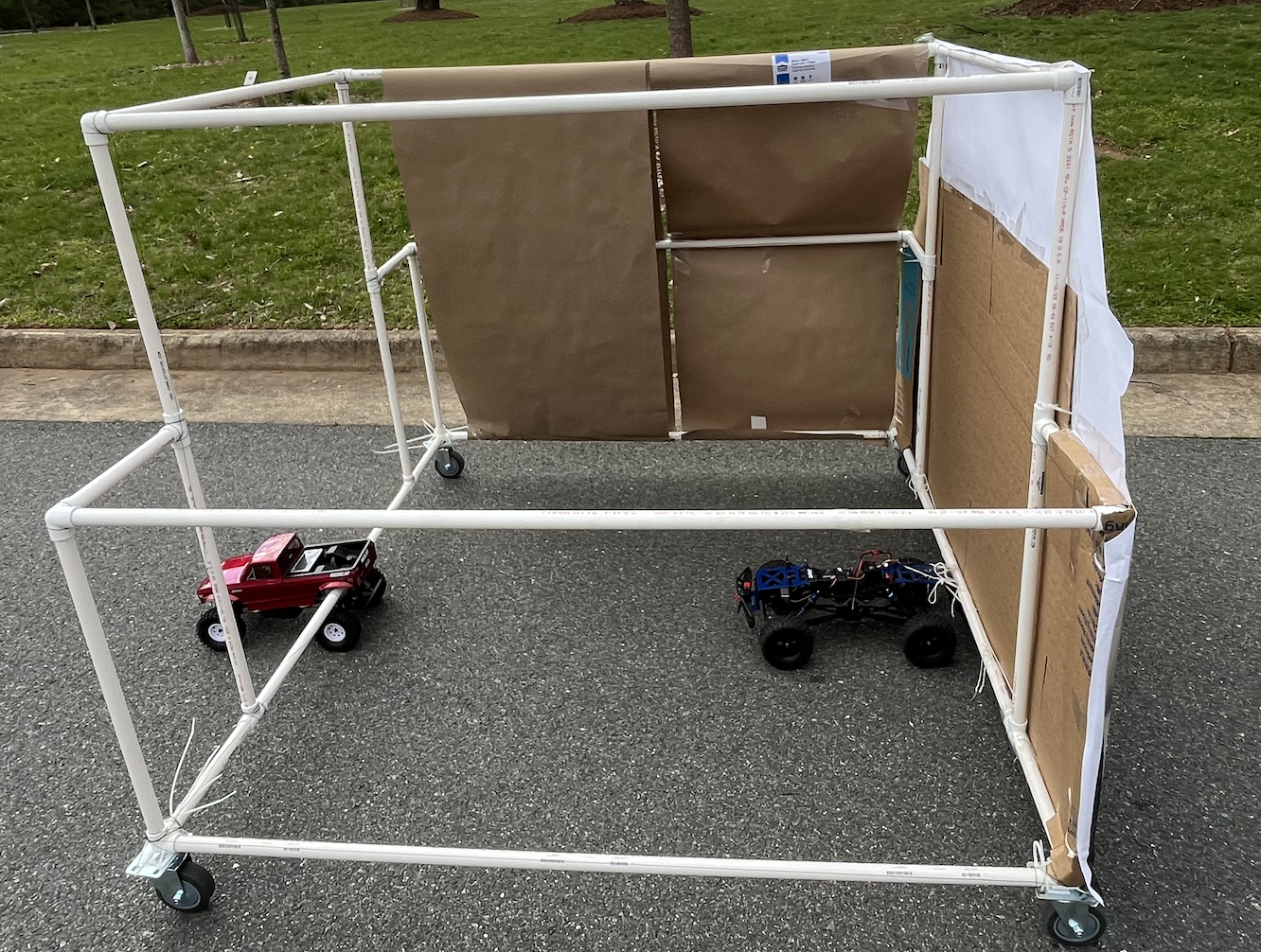}
        (2a)  
    \end{minipage}
    \begin{minipage}{0.3\columnwidth}
	\small 
        \centering
    \includegraphics[width=\columnwidth]{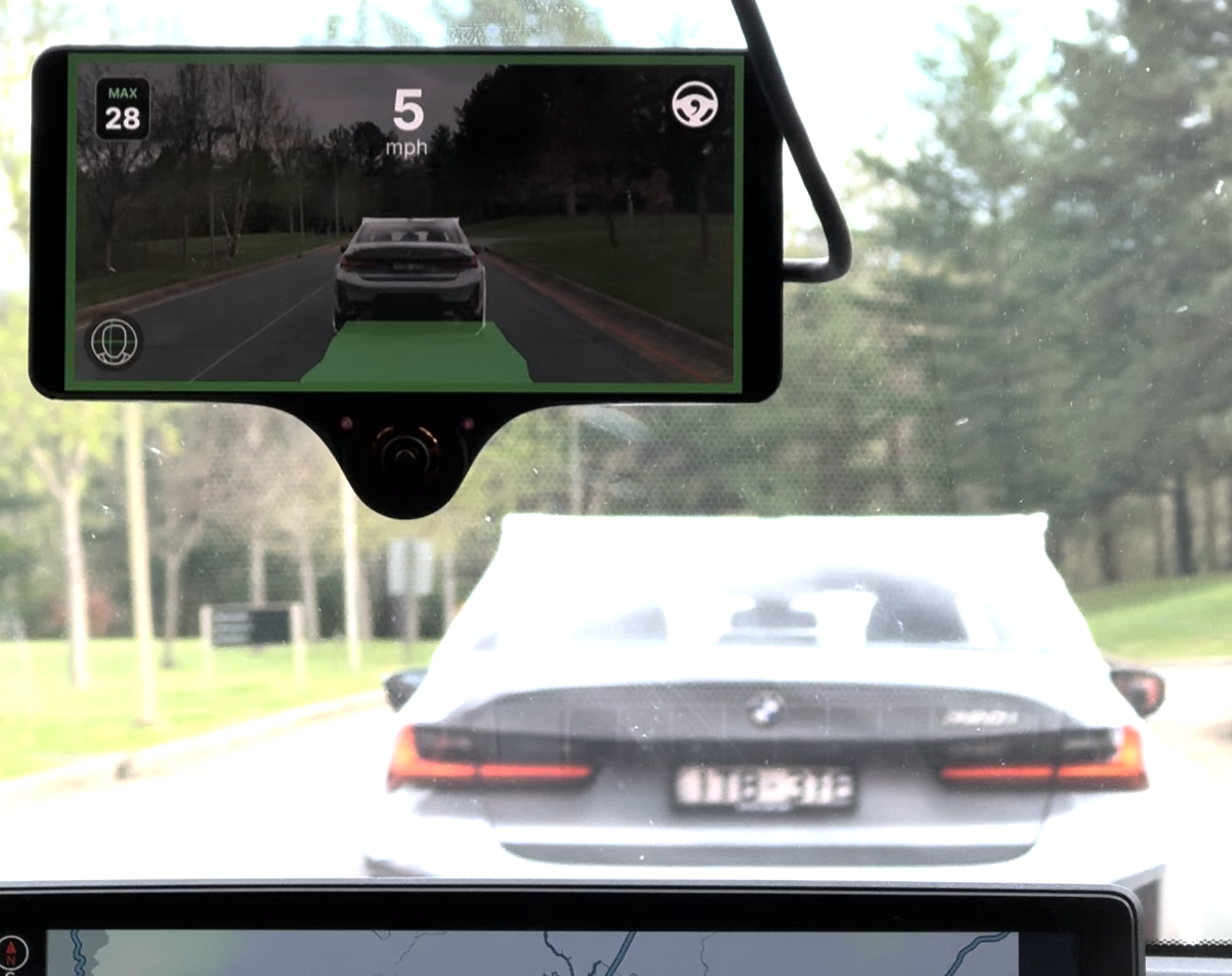}
    (2c)  
    \end{minipage}
    \begin{minipage}{0.3\columnwidth}
        \small
        \centering
        \includegraphics[width=\columnwidth]{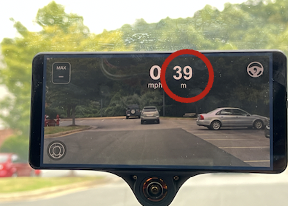}
        (1b)  
    \end{minipage}
    \hfil
    \begin{minipage}{0.3\columnwidth}
        \small
        \centering
        \includegraphics[width=\columnwidth]{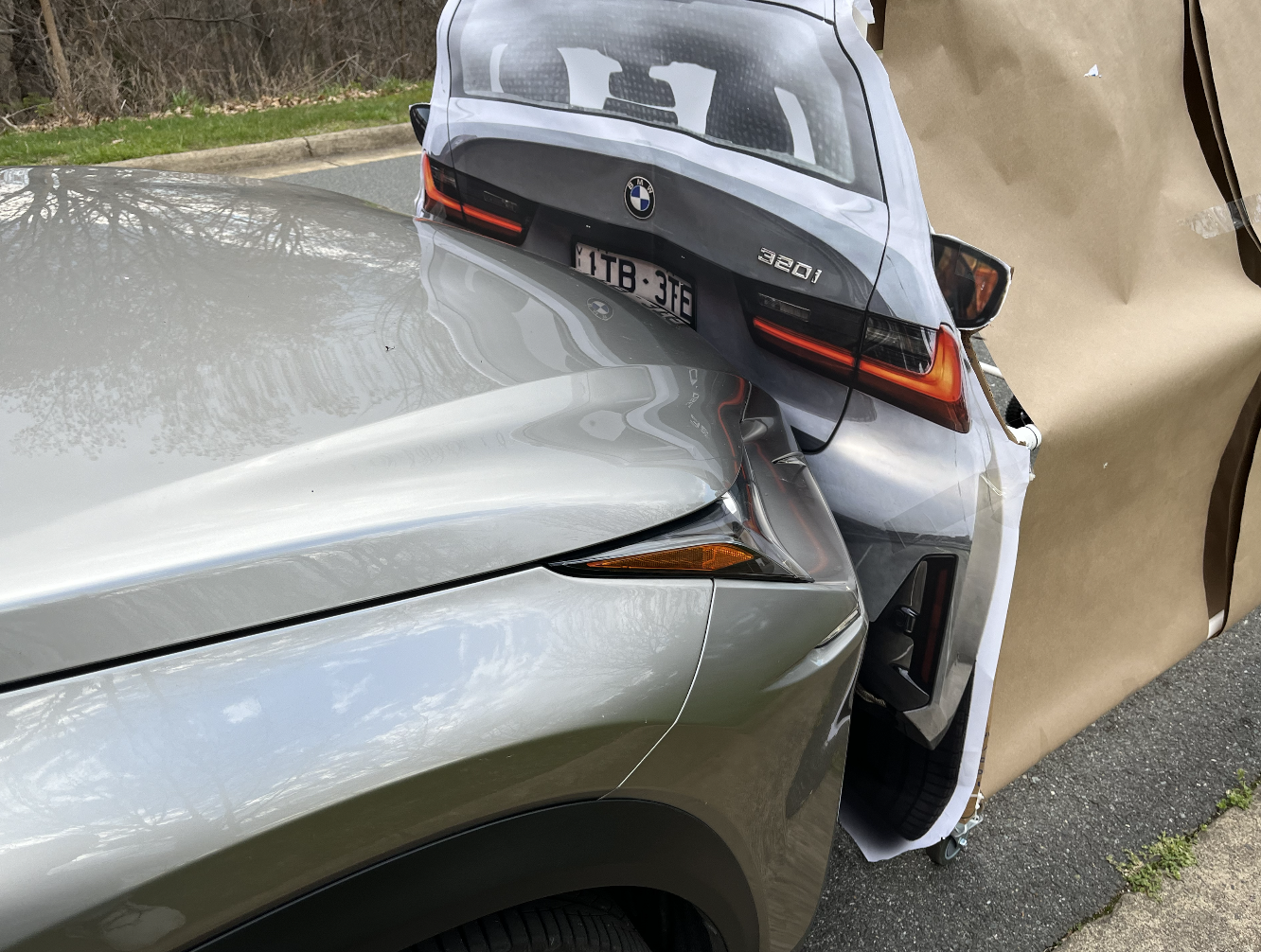}
        (2b)  
    \end{minipage}
    \begin{minipage}{0.3\columnwidth}
	\small 
        \centering
    \includegraphics[width=.9\columnwidth]{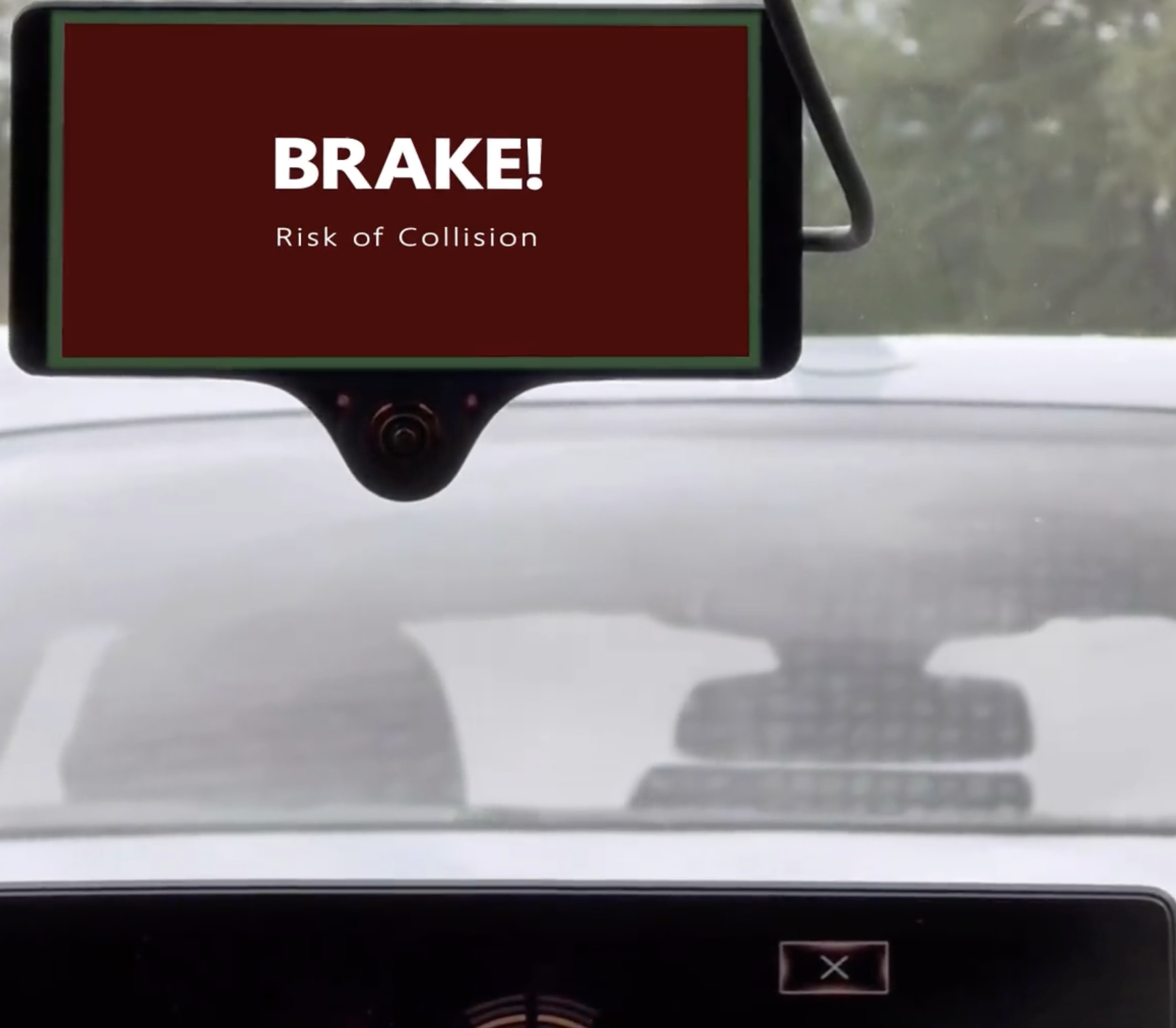}
    (2d)  
    \end{minipage}
    \vspace{-1em}
    \caption{RD predictions w/o (1a) or w (1b) patch on an actual vehicle in a parking lot; (2a) Side view of lead car model; (2b) AV under perception attack collides with the lead car model; (2c) AV follows the car model in a benign scenario; (2d) Driver's view upon collision. 
    } 
    \label{fig:carmodel}
    \label{fig:actualvehicle}
    \vspace{-2em}
\end{figure}

First, the Ego vehicle was parked at distances ranging from 10m to 50m (at intervals of 5m) to the LV. We modified the OpenPilot code to display the RD predictions on the device monitor, as shown in Fig. \ref{fig:actualvehicle}-1a,1b. In these tests, the CA-Opt attack caused an average deviation of 16.2m in distance predictions, which could likely lead to a forward collision in the end-to-end ACC. This conclusion is based on our simulation experiments, where deviations exceeding 10 meters triggered sudden accelerations, leading to forward collisions.


\begin{figure}[t]
    \centering
    \includegraphics[width=0.7\columnwidth]{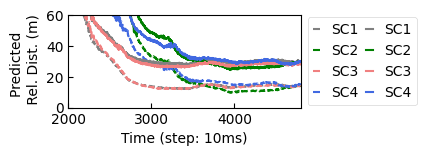}
    \vspace{-1.5em}
    \caption{Relative distance predictions with (solid lines) or w/o (dashed lines) adversarial patch for different driving scenarios.}
    \label{fig:cardistance}
    \vspace{-1em}
\end{figure}

Then, we conducted experiments using the same scenarios (SC1-SC4) outlined in Section \ref{sec:expsetup}. For an accurate assessment of the impact of the attack, we cloned the LVD's DNN model within the OpenPilot control software and ran both the original and the duplicate model on the AV simultaneously. During each perception cycle (20Hz), we initially supplied a benign image 
to the standard DNN model. Then, we duplicated this benign image, injected the adversarial patch to it, and then fed it to the second DNN model.  
The predictions from each model were recorded in separate log files. The results, presented in Fig. \ref{fig:cardistance}, indicate that the attack increased RD predictions by an average of 15.3 meters across all tested scenarios. 

We also conducted experiments using a real-world video dataset as described in Appendix \ref{appendix:dataset}.
{These experiments demonstrate the effectiveness of CA-Opt attack in impacting the DNN-based perception module in real-world driving scenarios.}


\textbf{End-to-End Evaluation.}
We also evaluate the impact of attacks on the end-to-end ACC on actual vehicle. To ensure the safety of both the driver and the vehicle, we constructed a lead car model from PVC pipe, designed to match the dimensions of a real BMW car model \cite{bmwsize}. We aimed for the OpenPilot system to recognize this fabricated car as a genuine vehicle by attaching a rear-view image of a car to its rear end (see Fig. \ref{fig:carmodel}-2c).  In this experiment, the AV approached the LV from a distance of 50 meters with a cruise speed set at 28 mph. Meanwhile, the LV was propelled by two remote-controlled ground robots (Fig. \ref{fig:carmodel}-2a). We conducted this experiment with and without attack (adversarial patches) activated. 

OpenPilot software successfully recognized the lead car model as a legitimate vehicle and maintained a safe following distance and speed (about 5 mph) in the benign scenario (see Fig. \ref{fig:carmodel}-2c). 
However, 
in the presence of the attack, the AV continued to advance toward the lead car and eventually collided with it (Fig. \ref{fig:carmodel}-2b), despite AEBS being activated (Fig. \ref{fig:carmodel}-2d). This underscores the generalization of our proposed attack in efficiently causing safety hazards and exposes the inadequacy of existing safety mechanisms in preventing the attack. Moreover, the time between the AEBS warning and the collision was approximately 1.5-2.2 seconds, shorter than the average driver reaction time of 2.5 seconds, leaving insufficient time for a human driver to intervene and prevent the collision.

\subsection{Evading Existing Defense Methods}
\label{sec:defense}
While our proposed stealthy adversarial patches are invisible to human eye, they may be detected by some existing defense methods.

\textbf{Adversarial Patch Detection.}
Methods such as gradient masking \cite{papernot2016towards}, lossy compression \cite{kurakin2018adversarial}, or adversarial training \cite{madry2017towards} have been proposed for adversarial patch detection. 
However, these methods either need to be trained on specific attacks with high computation costs \cite{xu_patchzero_2023,chen_jujutsu_2023,liu_segment_2022} or significantly sacrifice DNN prediction accuracy \cite{papernot2016towards,xiang_patchguard_nodate}, which negatively affect ACC systems' safety.

\begin{figure}[t]
    \centering
    \includegraphics[width=\columnwidth]{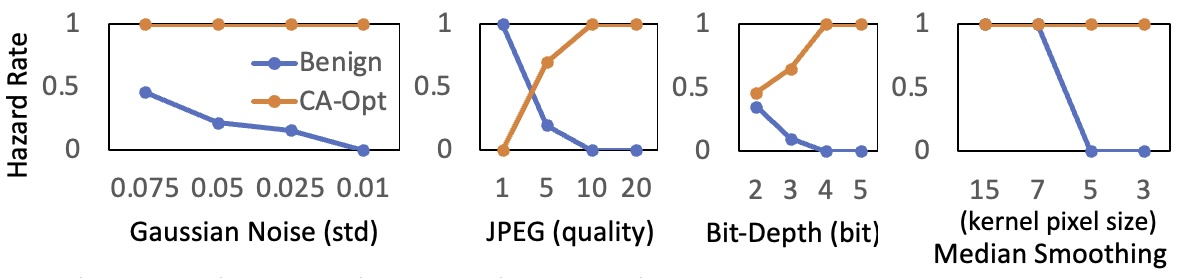}
    \vspace{-2.5em}
    \caption{Results of each directly-applicable defense method.}
    \label{fig:defense}
    \vspace{-2em}
\end{figure}

We assess four widely used open-source defense methods that only rely on model input transformation without the need for re-training, including 
adding Gaussian noise \cite{zhang2019defending}, JPEG compression \cite{dziugaite2016study}, reducing image color bit-depth \cite{xu2017feature}, and using spatial median smoothing \cite{xu2017feature}.
We implement all these defense methods by changing each input image frame with various parameter settings (as shown in the x-axis of Fig. \ref{fig:defense}). 
We evaluate the attack success rates in causing hazards under each defense method while considering the effect of input transformations on the benign or attack-free image frames to maintain the baseline safety of the ACC system. 

As shown in Fig. \ref{fig:defense}, JPEG compression and bit-depth reduction methods effectively reduce the attack success rate under specific parameter configurations. However, these methods fall short in maintaining the ACC's safety by leading the benign image frames to cause hazards. In instances where the benign cases do not lead to hazards, the attack hazard rate is at 100\%.
On the other hand, the incorporation of Gaussian noise or median smoothing reduces the ACC's LV detection accuracy. These methods are ineffective in mitigating CA-Opt attacks (hazard rate stays at 100\% for all configurations), while also causing hazards for benign frames.

These results indicate that our attack can easily evade the directly applicable defense methods. More advanced methods, such as adversarial training, may need to be developed/trained specifically for our design, which are subject to future direction.

\textbf{Sensor Fusion.}
An alternative defense against adversarial patches could involve integrating independent sensors like Lidar or radar with camera data for LVD predictions. However, Lidar is too costly for Level-2 AVs \cite{sato2021dirty}, and our tests found that radar-camera fusion did not prevent ACC misbehavior or collisions (see Appendix \ref{appendix:fusion}). This may be because of using Kalman filters for sensor fusion, which assume measurement noise is zero-mean Gaussian and are vulnerable to perturbations smaller than one standard deviation of this noise \cite{jha2020ml}. 
In addition, sensor fusion outputs a weighted summation of radar and camera predictions, which cannot completely eliminate deviations caused by erroneous camera predictions, particularly when they significantly deviate from the ground truth. In some production ACC, camera predictions carry more weight.  
The sensor fusion vulnerability was also reported in previous work~\cite{hallyburton2022security}.

\section{Discussion}
\label{sec:Discussion}

\textbf{Sim-to-Real Gap.}
Addressing the sim-to-real gap in AV security literature is challenging due to the risks and costs of real-road tests.
In this paper, we tried to narrow this gap by developing a realistic experimental platform that integrates production ADAS control software, a physical-world simulator, and well-designed safety interventions with high-risk driving scenarios designed based on the NHTSA report \cite{najm_pre-crash_2007}. Moreover, we evaluate the sim-to-real transfer possibility using an actual vehicle, a model lead car, and a publicly available dataset.  However, there are still some limitations, such as using fixed models and thresholds in the design of the driver reaction simulator, that may impact evaluation results. Also, additional firmware and safety checks might be deployed in real cars, which can further limit the effectiveness of the proposed attack.



\textbf{Attack Method Generalization.}
We demonstrate the generalization of our proposed attack on a production ACC system, OpenPilot, through closed-loop simulation, real-world AV dataset, and actual vehicle experiments. 
However, the vulnerability of other Level-2 production ACC systems, such as Tesla Autopilot or Cadillac Super Cruise, to our attacks remains uncertain due to their closed-source nature. While we cannot directly evaluate our attacks on these systems, it is reasonable to argue that our results hold generalization potential based on the representative nature of the OpenPilot ACC system.
Specifically, our attack strategy, which leverages context awareness derived from high-level system hazard analysis, can be generalized to diverse ACC systems. Furthermore, our optimization-based attack vector generation can be applied to other DNN-based ACC systems, given the inherent vulnerability of DNNs to adversarial input perturbations \cite{eykholt2018robust,sato2021dirty,jha2020ml,goodfellow2014explaining}.







\vspace{-0.5em}
\section{Related Work} 
\textbf{Adversarial Attacks on DNN.}
Many works have explored the vulnerability of DNN against adversarial attacks by adding adversarial physical/digital patches or stickers \cite{liu2018dpatch,lee2019physical,hoory2020dynamic,goodfellow2014explaining,moosavidezfooli2016deepfool,eykholt2018robust,sato2021dirty, tencent2019experimental,wu2020making}. 
However, most of these works focus on altering the prediction class or probability or lane line position, which do not apply to attacks against ACC. Moreover, they rely on off-line optimization of attack value, neglecting the impact of attack timing. In contrast, our work introduces a novel runtime perception attack method against DNN-based ACC systems, employing a combined knowledge and data-driven approach that considers both attack timing and value for enhanced effectiveness. 
The only other work on ACC \cite{guo2023adversarial} focused on the physical attacks without considering dynamic changes at runtime, which is not scalable to many vehicles.

\textbf{Security Analysis of AVs.}
Great efforts have also been made in studying the security of AVs, such as the security of Lidar \cite{lidar_2023}, GPS \cite{shen2020drift}, radar \cite{komissarov2021spoofing}, camera \cite{sato_wip_2023}, lane detection \cite{sato2021dirty, tencent2019experimental}, multiple objects tracking \cite{jha2020ml,jia2019fooling,ma_wip_2023}, control software \cite{zhou2022strategic,ding2021mini}, and safety mechanisms \cite{ma2021sequential}.
To the best of our knowledge, this paper is the first analysis of the security of Level-2 production ACC systems under stealthy safety-critical attack by considering three levels of safety interventions by constraint checking, human driver, and AEB/FCW and addressing unique challenges (Section \ref{sec:challenges}).

\section{Conclusion}
\label{sec:conclusion}

This paper proposes a novel runtime stealthy attack strategy against DNN-based ACC systems, consisting of (i) a control-theoretic method for finding the most critical system contexts for launching the attacks to maximize the chance of safety hazards and (ii) an optimization-based image perturbation method for efficient generation and injection of adversarial patches to the DNN input to cause ACC control misbehavior and hazards as soon as possible before being detected or mitigated by the ADAS safety mechanisms or human driver.  
Experiments on a production Level-2 ADAS using an enhanced closed-loop simulation platform, a publicly available driving dataset, and an actual vehicle demonstrate the effectiveness of our approach in improving attack success rate and stealthiness compared to different baselines. 
This study also provides insights into the development of future ADAS that are robust against safety-critical attacks and the importance of interventions by the drivers and basic safety mechanisms for preventing attacks.

\section*{Acknowledgment}
This work was partially supported by a gift from Toyota InfoTechnology Center and by the National Science Foundation (NSF) under Grants 2402941 and 1931997.
{
\bibliographystyle{unsrt}
\bibliography{main}
}

\appendix

\section{Sensor Fusion}
\label{appendix:fusion}


We implement a radar sensor in the CARLA simulator and feed the data to the OpenPilot radar interface \cite{weng2022formal}, to be used as an independent input by the fusion module. 
Specifically, we use the DBSCAN algorithm \cite{schubert2017dbscan} to cluster the 2D point map of the relative distance and speed of perceived objects from the radar sensor and feed their mean values to OpenPilot, which are then further filtered and processed for fusion.

\begin{figure}[h!]
    \centering
    \vspace{-0.5em}
    \includegraphics[width=0.9\columnwidth]{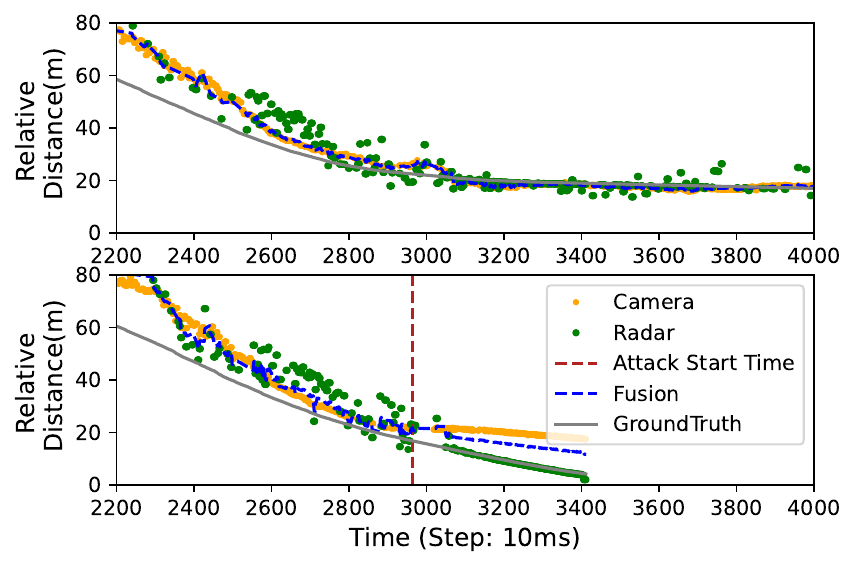}
    \vspace{-1.5em}
    \caption{An example fusion of relative distance predictions based on camera and radar data compared with the ground truth under normal operation (Top) or under attack (Bottom).}
    \label{fig:fusionexample}
    \label{fig:fusion-attack}
    \vspace{-1.5em}
\end{figure}

Fig. \ref{fig:fusionexample} (Top) shows an example of predictions of the relative distance to the lead vehicle from the fusion of the camera and radar measurements. We see that the radar and camera predictions agree well most of the time. 
Also, the error between the fusion predictions and the ground truth relative distance (based on positions of vehicles in the simulator) becomes smaller as the Ego vehicle approaches the lead vehicle (an RMSE of 0.81m after 3,000 control cycles in the figure). 
Sensor fusion also helps reduce the errors in fusion predictions under attacks as shown in Fig. \ref{fig:fusion-attack} (Bottom) and discussed in Section \ref{sec:defense}, even though it fails to prevent collisions in the end.

\section{Malware Installation}
\label{Appendix:deployment}
\begin{figure}[h!]
    \centering
    \vspace{-0.5em}
    \includegraphics[width=\columnwidth]{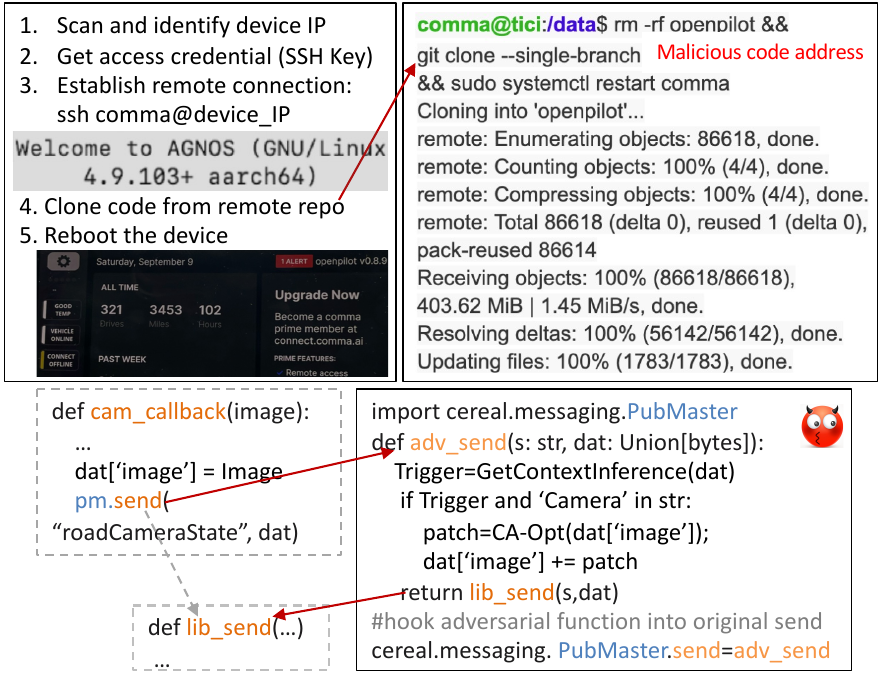}
    \vspace{-2em}
    \caption{An example of malware installation (Top) and malicious code execution using monkey patching technique (Bottom).}
    \vspace{-1.5em}
    \label{fig:malwaredeployment}
\end{figure}

Fig. \ref{fig:malwaredeployment} shows an example set of steps taken for establishing remote access and downloading a malicious code repository on a vehicle running OpenPilot, the open-source production ACC system from Comma.ai \cite{commaai-openpilot}.
To change live camera image frames at runtime, we use a technique called monkey patching in Python, which is used for dynamically modifying or extending the behavior of an existing code at runtime or hooking a function without changing the source code\cite{hunt2019monkey}.
For example, as shown in Fig. \ref{fig:malwaredeployment}, the \textit{send()} system library called by the camera call back function for sending the received camera frames to the DNN model can be wrapped by a malicious version \textit{adv\_send()} that implements the attack.

\section{AEBS Evaluation}
\label{appendix:survey-aebs}


For a realistic implementation of AEBS in our simulation platform, we study the AEBS design of typical OpenPilot-supported car models \cite{Carmodels-OpenPilot}. 
Our AEBS design relies on the fusion of camera and radar data, as mentioned in Appendix \ref{appendix:fusion}, aligning with AEBS design of most current OpenPilot-supported car models, as shown in Table \ref{tab:survey-aebs}.

\begin{table}[t]
\footnotesize
\centering

\caption{AEBS design in OpenPilot-supported car models.} 
\vspace{-1.5em}
\label{tab:survey-aebs}
\resizebox{\columnwidth}{!}
{%
\small
\begin{tabular}{@{}lllll@{}}
\toprule
Car Model  \cite{Carmodels-OpenPilot} & Is AEBS using Radar/Camera/Both \\ \midrule
 Acura RDX 2018& Both \\
 Buick LaCrosse 2019& Camera or both \\
 Cadillac Escalade 2017 & Radar or camera and ultrasonic sensors \\
 GMC Acadia 2018 & Camera and/or radar \\
 Honda Pilot 2022 & Both \\
 Honda Ridgeline 2023 & Both \\
 Lexus ES Hybrid 2023 & Both \\
 Lexus IS 2023 & Both \\
 Toyota Avalon Hybrid 2022 & Both \\
 Toyota Camry 2023 & Both \\
\bottomrule
\end{tabular}%
}
\end{table}

\label{appendix:aeb}

Following the testing protocol specified in \cite{http://data.europa.eu/eli/reg/2020/1597/oj}, we employ two categories and five driving scenarios (see Table \ref{tab:aebs-sce}) to assess our AEBS functionality (Section \ref{sec:AEB}). Each scenario is repeated 100 times to ensure reliable outcomes.
Experimental results show that in all five testing scenarios, both FCW and AEB alerts are activated, effectively preventing all hazards or collisions. On average, it takes about 1.68 seconds for AEB to stop the Ego vehicle completely.

\begin{table}[t]
\centering

\caption{Driving scenarios to test the AEBS with different initial distances ($Init\_dist$) between the Ego vehicle and the lead vehicle.} 
\vspace{-1em}
\label{tab:aebs-sce}
{%

\begin{tabular}{@{}lllll@{}}
\toprule
Lead vehicle & $Init\_dist$(m) & $V_{Ego}$(km/h) & $V_{Lead}$ (km/h) \\ \midrule
Stationary & 100, 100, 150 & 20, 42, 58 & 0\\ 
Moving & 100, 150 & 30, 58 & 20\\
\bottomrule
\end{tabular}%
}
\vspace{-1.5em}
\end{table}


{

\section{Stealthiness User Study}
\label{sec:appendix:userstudy}
We conduct a user study \cite{adassurvey} to further evaluate the advantages of the stealthiness design of our attack. Before recruiting participants, we secured Institutional Review Board (IRB) approval. Our study explicitly avoided collecting any personally identifying information, targeting sensitive populations, or introducing any risks to the participants.
Our study included 30 participants who were asked to sit on the driver's side of an autonomous vehicle, parked in a parking lot, equipped with OpenPilot ADAS (see Section \ref{sec:background:platform}). Each participant went through different trials of pre-recorded videos displayed on the ADAS monitor and answered a series of questions after each trial using a Qualtrics survey~\cite{qualtrics}. All participants had driving experience, and 40\% of them had AV driving experience. 

At the beginning of the study, we provide an introduction of ADAS and present demo videos on the ADAS monitor to ensure that the participants fully understand what driving technology we are surveying.  

\textbf{Driving Preferences. }We first ask participants to envision themselves driving this autonomous vehicle with the ADAS monitor displaying pre-recorded image frames. We inquire about how often they would look at the ADAS monitor while driving and whether alterations in the monitor's position and size influence their preference. User study results in Fig. \ref{fig:studyres1}
show that 99\% of the participants prefer looking at the ADAS monitor during their driving experience, with 33\% specifying they would do so for the majority of the driving duration. Moreover, 60\% of the participants indicate a preference for a larger monitor size or a more prominent position.

\textit{These results indicate that the driver might notice the camera input attacks and stealthiness design might be beneficial for these attacks to evade driver intervention.}

\begin{figure}[t]
    \centering
    \begin{minipage}[t]{0.49\columnwidth}
\includegraphics[width=\textwidth]{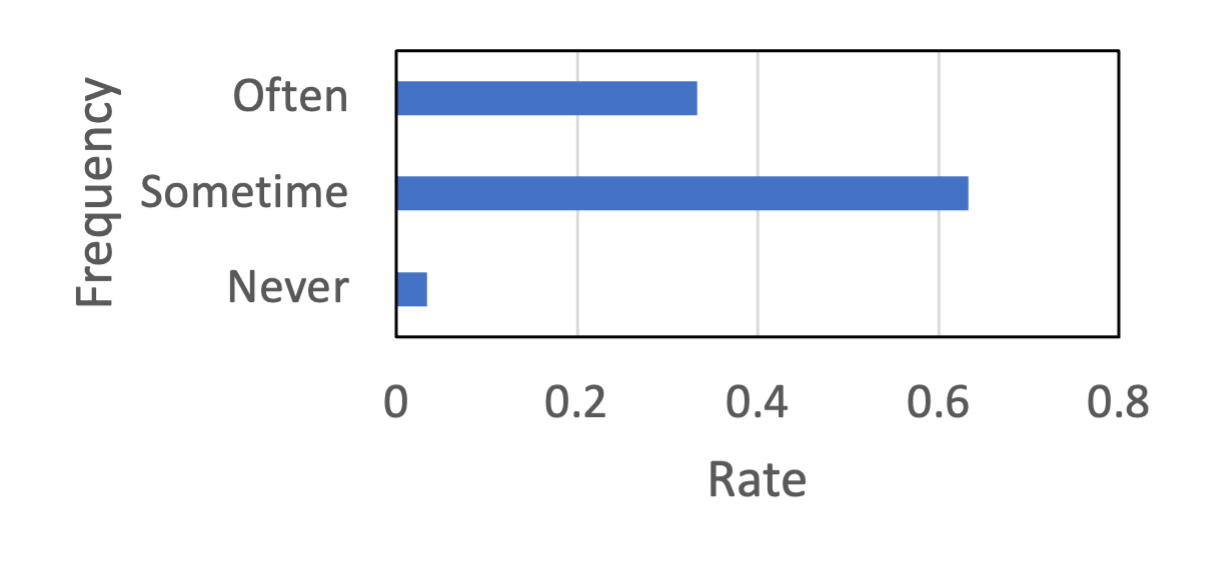}
\end{minipage}
\begin{minipage}[t]{0.49\columnwidth}
\includegraphics[width=\textwidth]{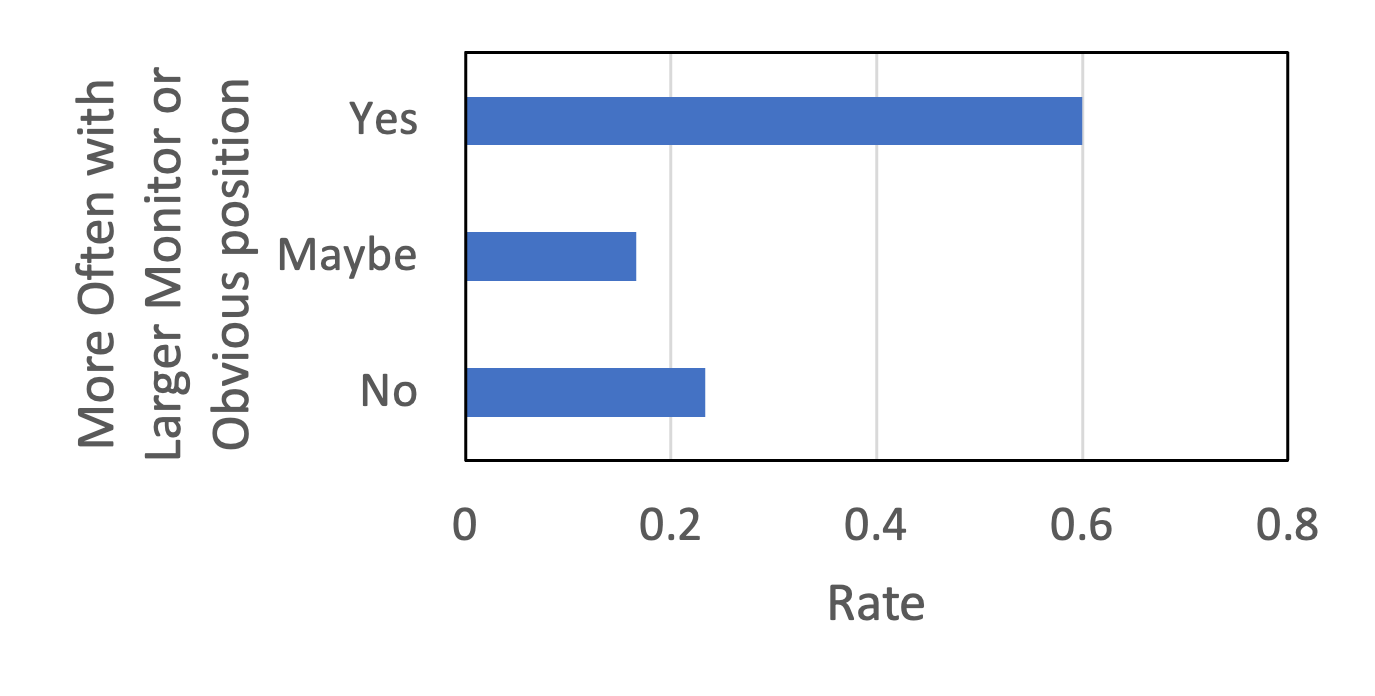}
\end{minipage}
    \vspace{-1.5em}
    \caption{Results of participants' preference of looking at the ADAS monitor during driving and whether they would look more often at the monitor with a larger size or in a noticeable position.}
    \vspace{-1em}
    \label{fig:studyres1}
\end{figure}

\textbf{Stealthiness. }We create five video sets by introducing adversarial patches into a pre-recorded highway scenario using CA-Random, CA-APGD, and CA-Opt methods with three stealthiness levels ($\lambda=10^{-4}$, $\lambda=10^{-3}$, $\lambda=10^{-2}$, as detailed in Section \ref{sec:evaluation:stealthuness}). We present these videos on the ADAS monitor and ask participants whether they notice any abnormal scenarios that prompt them to assume control of the vehicle to avoid potential risk or danger. For a detailed examination, we extract an image frame from each video at the same frame index and zoom in to reveal more intricate details, followed by posing identical questions to the participants.

\begin{figure}[h]
    \centering
    \vspace{-1em}
    \includegraphics[width=0.8\columnwidth]{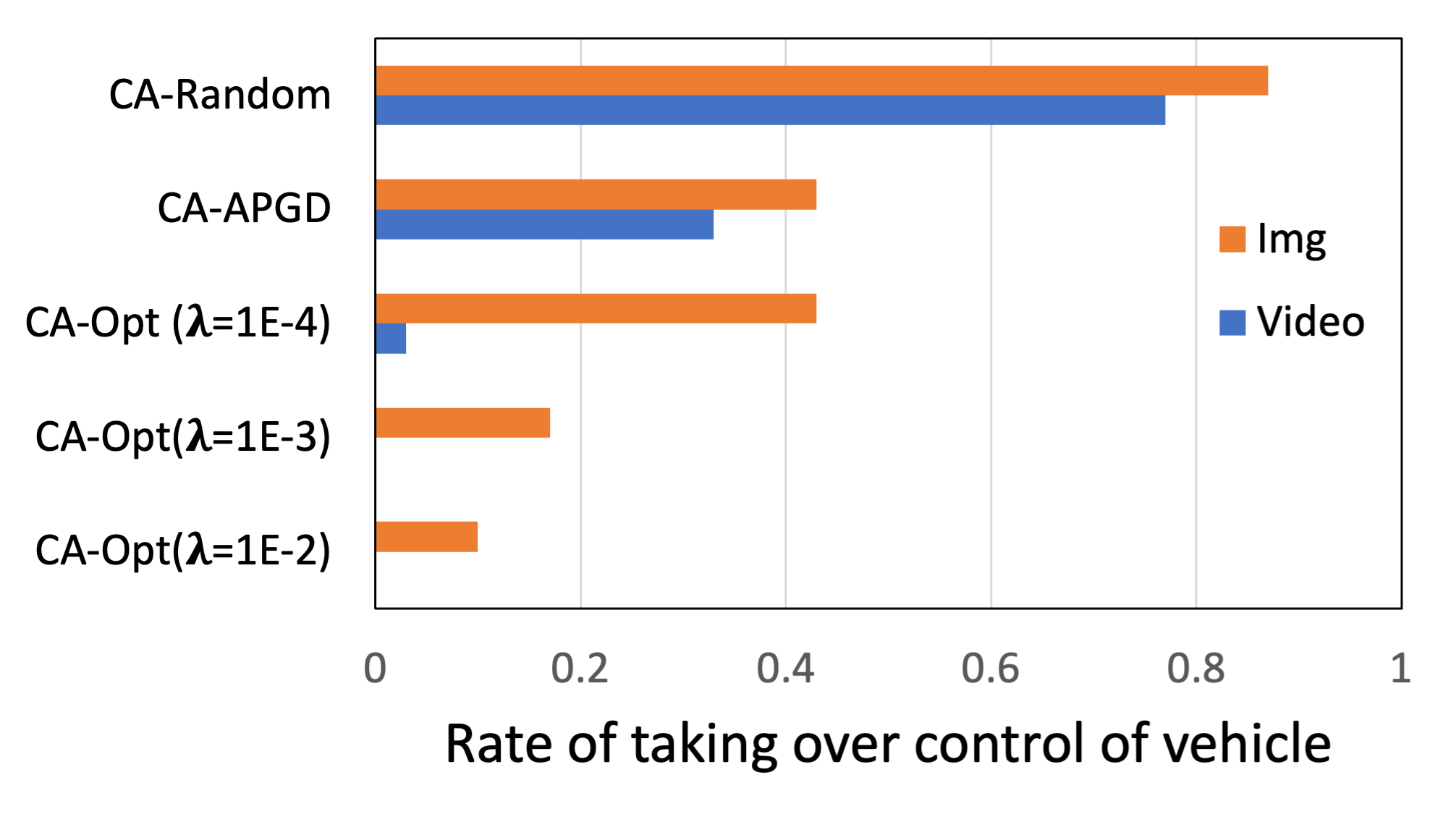}
    \vspace{-1.5em}
    \caption{Results of stealthiness of each attack method.}
    \vspace{-1em}
    \label{fig:studyresult4}
\end{figure}

User study results are illustrated in Fig. \ref{fig:studyresult4}. It is evident that patches generated by the CA-Random attack are conspicuous to the majority of participants (>75\%), whether observed in images or videos. In comparison, patches generated by the CA-APGD exhibit lower visibility than those produced by the CA-Random attacks.
In CA-Opt attacks, the takeover rate diminishes with a rise in stealthiness level or $\lambda$ value. Specifically, when $\lambda$ is set at $10^{-2}$ and $10^{-3}$, the takeover rates are below 20\% for patch images and are 0\% for patch videos. These findings suggest adversarial patches at $\lambda$ = $10^{-2}$ and $\lambda=10^{-3}$ exhibit nearly imperceptible characteristics to human drivers, particularly in image frames when not zoomed in.

\textbf{Physical Attack.} 
We also investigate the stealthiness of physical adversarial patches as perceived by human eyes. Participants are shown an image of an adversarial patch generated through a physical attack method introduced in a prior work \cite{guo2023adversarial}. They are then asked identical questions. Our findings reveal a takeover rate of 80\%, highlighting the inadequacy of physical patches in achieving stealthiness and evading human detection. 

\section{Evaluation of Fake Video Attacks}
\label{sec:appendix:fakevideo}
To further evaluate the necessity of a stealthy patch attack, we conduct another camera attack experiment by fake video injection. 

\textbf{Video Recording.} An Ego vehicle is configured to cruise at 40mph from 75 meters away behind a lead vehicle cruising at 35 mph in CARLA simulator. We record the image frames captured by the camera on the Ego vehicle with a duration of 50 seconds. We select a portion of the recorded image frames (7 seconds) within a straight road area to be injected at runtime.

\textbf{Fake Video Attack.} We rerun the simulations for each scenario introduced in Section \ref{sec:expsetup} and replace the real-time camera frames with the selected fake video when the Ego vehicle approaches a similar position indicated by the fake video. Experimental results show that this attack causes hazards in 100\% simulations and in 72.6\% of simulations the Ego vehicle drives to the neighbor lane without any collisions. This is because lane lines in the fake video do not exactly overlap with the ones in the actual video. 

Therefore, we compare the recorded video to the real-time image frame captured by the Ego vehicle under attacks frame by frame and select the attack start time such that the fake image frame almost matches the real-time image frame (note that this selection of perfect match at runtime attack might be impossible). We rerun the simulations and experimental results show that the perfect fake video attack achieves a success rate of 95.1\% in colliding with the lead vehicle or side objects (e.g., road guard). 
The lower success rate of fake video attacks compared to the CA-Opt attack (100\%, as shown in Fig. \ref{fig:resOpt}) might be due to the difference between attack start times. 
We do not apply the context-aware strategy to fake video attacks since it determines the attack start time dynamically at runtime, and it is challenging to select a fake image frame that perfectly matches the image frame at the time inferred by context-aware strategy at runtime. 

\textbf{Observation.} 
Implementing a stealthy fake video attack is challenging as the attacker does not know the lanes the Ego vehicle will drive in the future, the positions and colors of surrounding vehicles, or the weather and road conditions. So, these differences between the fake video and the actual environment might trigger safety interventions and lead to mitigation of the attack. 
%

\begin{figure}[b]
    \centering
    \vspace{-1em}
    \begin{minipage}[t]{0.35\columnwidth}
\includegraphics[width=\textwidth]{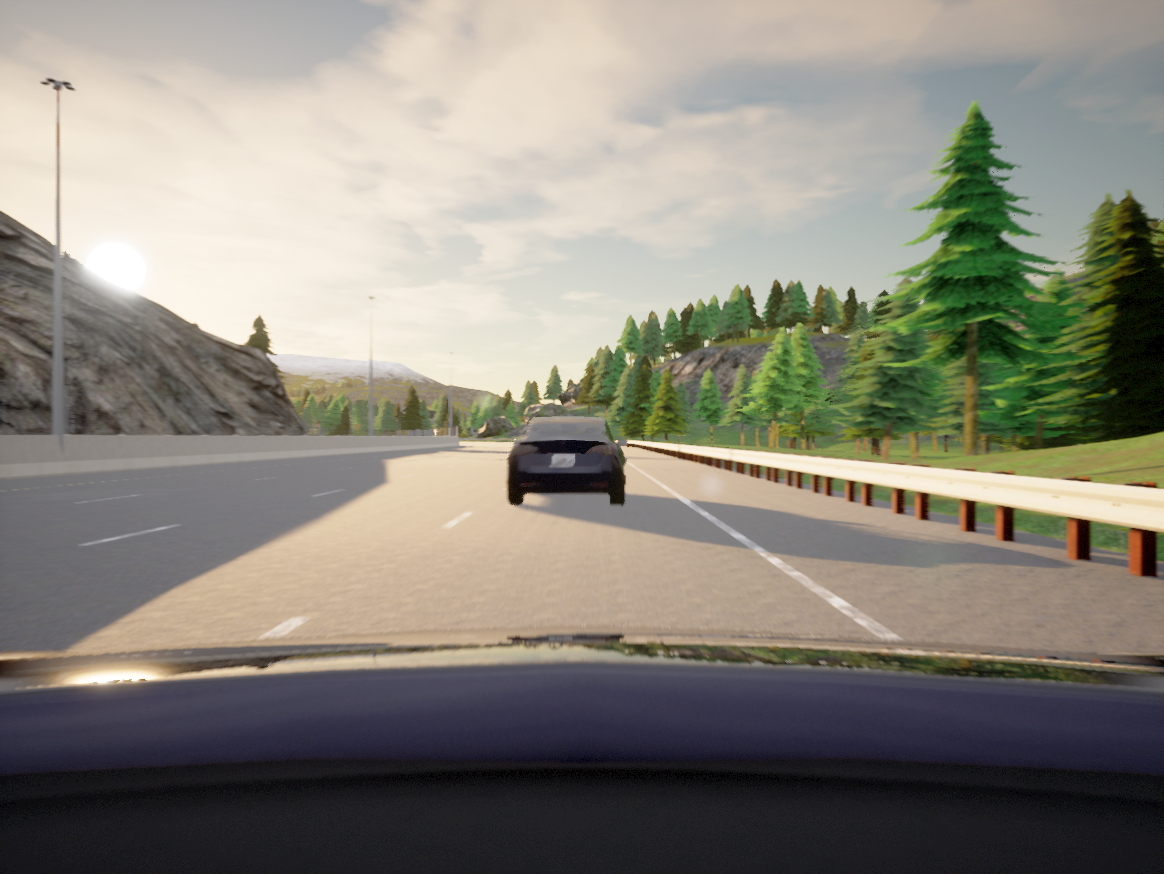}
\end{minipage}
\begin{minipage}[t]{0.35\columnwidth}
\includegraphics[width=\textwidth]{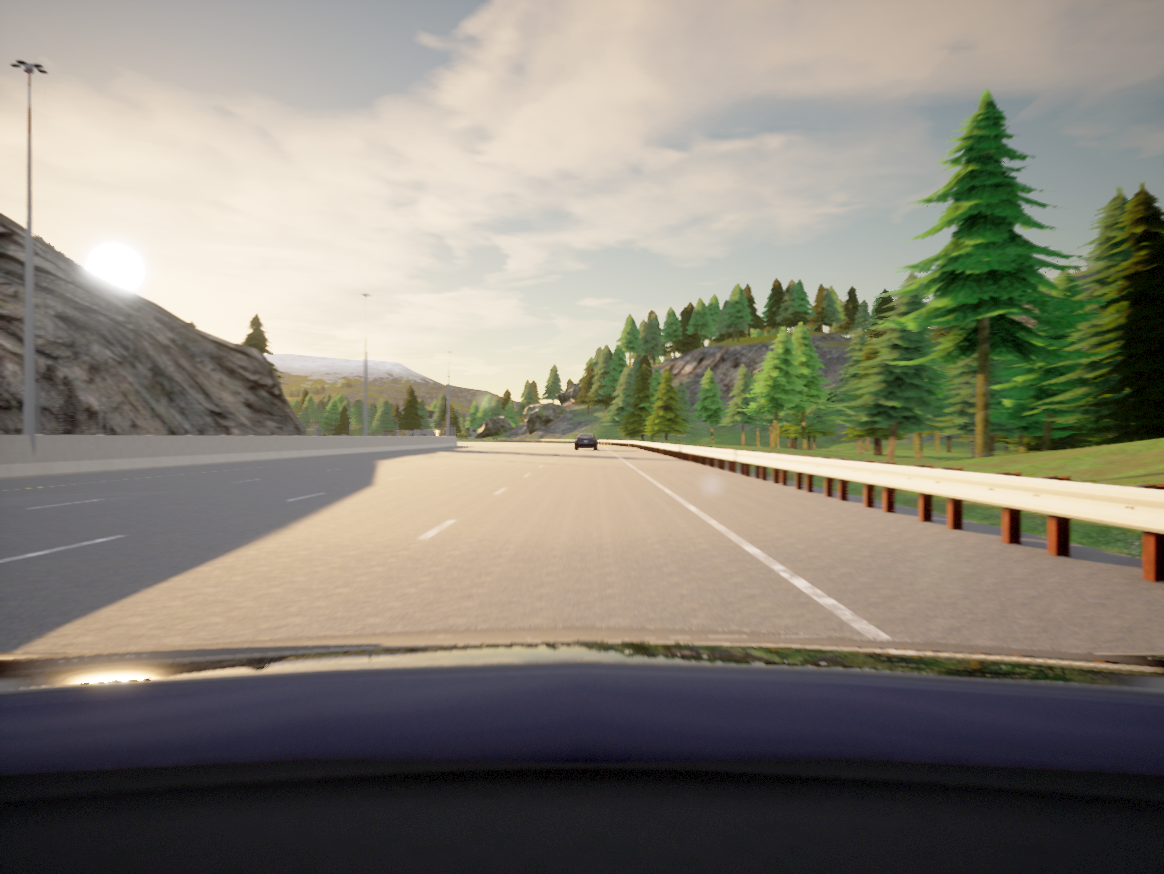}
\end{minipage}
    \vspace{-1em}
    \caption{An example of two consecutive images before (Left) and after (Right) fake video attack.}
    
    \label{fig:fakeimgs}
\end{figure}

Due to such differences, fake video attacks can be easily detected by existing methods that monitor the differences between two consecutive image frames. An example of image frame changes during a perfect fake video attack is shown in Fig. \ref{fig:fakeimgs}. Even if the fake videos are recorded using the same camera on the Ego vehicle driving in the same lane and weather conditions with only one lead vehicle (resembling replay attacks), an alert human driver can still notice the changes in the lead vehicle's position and size. An example of the RMSE and UIQ \cite{wang2002universal} between two consecutive image frames is shown in Fig. \ref{fig:fakeimgRmse}. We see that the similarity between the first frame of the fake video and the last frame of the benign video is much lower than that between other consecutive frames. 

}

\begin{figure}
    \centering
    \begin{minipage}[t]{0.49\columnwidth}
        \includegraphics[width=\columnwidth]{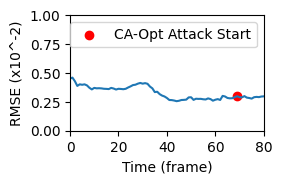}
    \end{minipage}
    \begin{minipage}[t]{0.47\columnwidth}
        \includegraphics[width=\columnwidth]{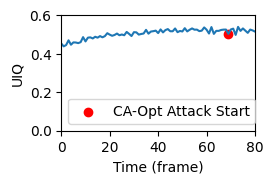}
    \end{minipage}
    \begin{minipage}[t]{0.49\columnwidth}
        \includegraphics[width=\columnwidth]{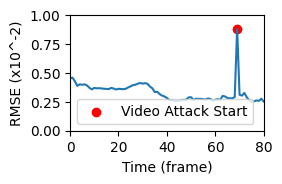}
    \end{minipage}
    \begin{minipage}[t]{0.47\columnwidth}
        \includegraphics[width=\columnwidth]{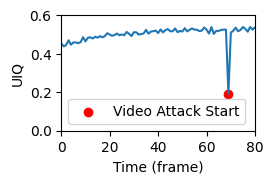}
    \end{minipage}
    \vspace{-1em}
    \caption{Similarity of two consecutive image frames with CA-Opt attack and fake video attack (starting at 69th frame) measured in RMSE and universal image quality index (UIQ) \cite{wang2002universal}.}
    \vspace{-1em}
    \label{fig:fakeimgRmse}
\end{figure}

{
\textbf{Fake Video Attack with Safety Interventions.} To further evaluate the performance of fake video attacks with safety mechanisms, we rerun the experiments by launching the proposed driver intervention 2.5 seconds (average reaction time) after the attack. 
Experimental results show that all the attacks are successfully prevented. Therefore, we do not further test the fake video attack while enabling AEBS and constraint checking (see Section \ref{sec:platform}).
}

\section{Robustness to Real-world Factors}
\label{sec:robustness}
\begin{figure}[b]
    \centering
    \includegraphics[width=0.7\columnwidth]{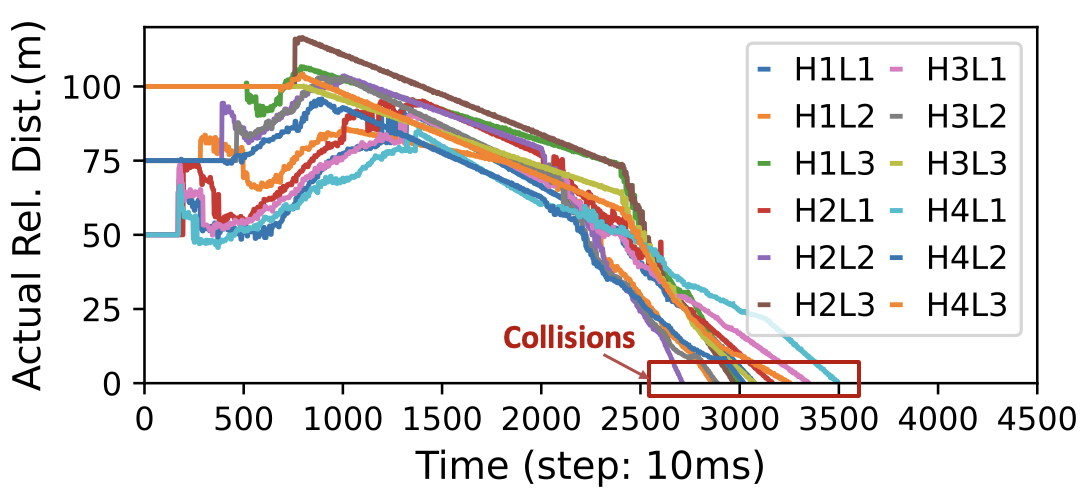}
    \vspace{-1em}
    \caption{Actual relative distance trajectories under CA-Opt attack with different camera heights (H1:1.1m, H2:1.3m, H3:1.5m, H4:1.7m) and initial longitudinal distances to the lead vehicle (L1:50m, L2:75m, L3:100m). An actual relative distance of zero indicates collision.}
    \label{fig:robust}
\end{figure}

To assess attack robustness, we vary front camera height based on standard passenger car profiles from manufacturers \cite{urazghildiiev2007vehicle}. We perform our experiments with four heights between 1.1-1.7 meters and three initial distances (50m, 75m, 100m). 
Fig. \ref{fig:robust} illustrates the 100\% success rate of our CA-Opt attack across 12 testing scenarios. The Ego vehicle initially maintains a safe following distance, deviates from it around the 2,500-3,000 control cycle or step due to the adversarial patch, and eventually collides with the lead vehicle. These results demonstrate our attack is robust to different camera positions and initial longitudinal distances and can cause safety hazards. 
We also test our attacks with diverse weather (rainy, sunny, or cloudy) and lighting conditions (noon or sunset). Results show that our CA-Opt attack causes longitudinal deviations of 9.8-14.3m in the predicted lead vehicle position, while maintaining a success rate of 100\% under such conditions. 


\section{Runtime Overhead}
\label{sec:timeliness}
\begin{figure}[t!]
    \centering
    \includegraphics[width=.9\columnwidth]{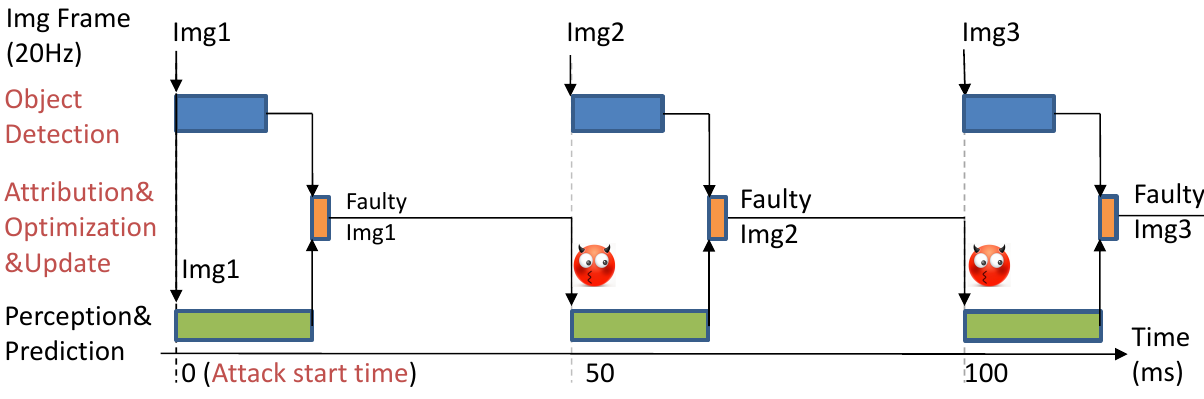}
    \vspace{-0.5em}
    \caption{Runtime overhead of each step of the attack.}
    \vspace{-1em}
    \label{fig:timeoverhead}
\end{figure}

To further assess our attack's real-world applicability, we measured its runtime overhead on a Comma3 device. We parked the Ego vehicle, equipped with OpenPilot and our attack malware, behind a lead vehicle in a parking lot, activating the ACC function with a cruise speed set to 0 mph. We record the time overhead for each component, as shown in Fig. \ref{fig:timeoverhead}, and report the average value over 5,000 control cycles. 

Experimental results show that the time overhead introduced by the context inference component before activating the attacks is minimal (1.17 us). 
Following the activation of attacks, the time overhead for the object detection module is about 10.1 ms on average. 
Note that some production ADAS provide object detection and tracking features, so this overhead time could be potentially avoided. 
We also observe that the primary attribution algorithm \cite{sundararajan2017axiomatic} does not add significant overhead, leveraging gradients calculated during the patch optimization process. 
The total time overhead is 1.52 ms.


\section{Testing on a Real-world Dataset.}
\label{appendix:dataset}
We perform an evaluation using the comma2k19 dataset \cite{comma2k19}, a publicly available dataset with over 33 hours of California's 280 highway commute.  The dataset comprises 2019 segments, each lasting one minute, covering a 20km highway section, collected using OpenPilot hardware. 
From this dataset, we choose 200 videos with a clear view of the lead vehicle and a relative distance of less than 100 meters. These videos are fed into the DNN model to record predictions for relative distance (considered as ground truth). We then introduce adversarial patches, generated by different attack methods, into the videos. The manipulated videos are fed into the DNN model, and predictions for relative distance are compared with those from videos without any attacks, using metrics like average and standard deviations in the predicted longitudinal distance.


\begin{table}[b]
\vspace{-1.5em}
\caption{Performance in deviating DNN-based lead vehicle position predictions using comma2k19 dataset.}
\vspace{-1em}
\label{tab:res2k19}
\resizebox{\columnwidth}{!}{%
\begin{tabular}{@{}llllllll@{}}
\toprule
\multirow{2}{*}{\textbf{Attack}} &\multirow{2}{*}{\textbf{Metric}} & \multicolumn{6}{c}{\textbf{Longitudinal Deviation in DNN Prediction }(m)} \\ \cmidrule(l){3-8}
& & 0-20 & 20-40 & 40-60 & 60-80 & 80+ & All \\ \midrule
\multirow{2}{*}{CA-Random} & Avg.  & 0.70 & 0.41 & 0.28 & 0.99 & 0.08 & 0.15 \\
& Std & 0.69 & 1.03 & 1.44 & 1.93 & 2.67 & 2.56 \\\midrule


\multirow{2}{*}{CA-Opt} & Avg. & 18.65 & 16.15 & 14.52 & 8.65 & 3.73 & 4.91 \\
& Std & 6.96 & 4.83 & 4.95 & 2.82& 3.03 & 3.35\\

\bottomrule
\end{tabular}%
}
\end{table}

Table \ref{tab:res2k19} compares the CA-Opt attack with CA-Random for different distances between the Ego and lead vehicles. CA-Random has an average deviation of 0.15m, which does not significantly impact ACC system outputs or cause hazards as the ACC system typically keeps a following distance larger than 4 meters in the absence of attacks. 
In contrast, when the Ego vehicle is close to the lead vehicle (less than 20m), CA-Opt achieves the highest deviation of 18.65m, showcasing the effectiveness of the proposed objective function (Section \ref{subsec:objectivefunction}) in generating optimal perturbations with substantial impact. 



\end{document}